%% file: main.tex
\documentclass[ALICE,manyauthors]{cernphprep}
\usepackage[comma,square,numbers,sort&compress]{natbib}

\usepackage{lineno}
\usepackage{xspace}
\usepackage{hyperref}
\usepackage{multirow}
\usepackage{tabularx}
\usepackage{braket}
\usepackage{bm}
\usepackage{amsmath}
\usepackage[T1]{fontenc}
\usepackage{orcidlink}

\begin{document}
\input{commands.tex}

%%%%%%%%%%%%%%%  Title page %%%%%%%%%%%%%%%%%%%%%%%%
\begin{titlepage}
% the dates below correspond to CERN approval
% please don't touch: EB chairs will take care
\PHyear{2026}       % required, will be obtained from CERN
\PHnumber{117}      % required, will be obtained from CERN
\PHdate{15 April}  % required, will be obtained from CERN
%%%%%%%%%%%%%%%%%%%%%%%%%%%%%%%%%%%%%%%%%%%%%%%%%%%%

%%% Put your own title + short title here:
\title{$\mathbf{\ensuremath{{\textbf{K}}^{0}_{\textbf{S}}}\xspace}$--$\mathbf{\ensuremath{{\textbf{K}}^{0}_{\textbf{S}}}\xspace}$ femtoscopy in \mbox{Pb--Pb}\xspace collisions at $\mathbf{\sqrt{\textbf{\textit{s}}_{\textbf{NN}}} = 5.02~\textbf{TeV}}$ at the LHC}

\ShortTitle{\ensuremath{{\rm K}^{0}_{\rm{S}}}\xspace--\ensuremath{{\rm K}^{0}_{\rm{S}}}\xspace femtoscopy in \mbox{Pb--Pb}\xspace collisions at $\sqrt{s_{\mathrm{NN}}}~=~5.02$~Te\kern-.1emV\xspace}   % appears on left page headers

%%% Do not change the next lines
\Collaboration{ALICE Collaboration\thanks{See Appendix~\ref{app:collab} for the list of collaboration members}}
\ShortAuthor{ALICE Collaboration} % appears on right page headers, do not change

\begin{abstract}
Results from a one-dimensional femtoscopic analysis of \ensuremath{{\rm K}^{0}_{\rm{S}}}\xspace--\ensuremath{{\rm K}^{0}_{\rm{S}}}\xspace correlations in \mbox{Pb--Pb}\xspace collisions at the center-of-mass energy $\sqrt{s_{\mathrm{NN}}}~=~5.02$~Te\kern-.1emV\xspace using data collected by the ALICE experiment at the LHC are presented. 
The source radius $R$ and correlation strength $\lambda$ are studied as a function of centrality and 
pair-transverse momentum (\ensuremath{k_{\rm T}}\xspace) to provide insight into the space-time structure and composition of the particle-emitting source.
The observed trends of radii as a function of \ensuremath{k_{\rm T}}\xspace and centrality are 
consistent with the collective expansion of the system.
Comparisons with measurements at $\sqrt{s_{\mathrm{NN}}}~=~2.76$~Te\kern-.1emV\xspace by the ALICE Collaboration show agreement across multiplicities and \ensuremath{k_{\rm T}}\xspace.
Hydrokinetic model predictions match the most central collision results but deviate in peripheral events, 
potentially reflecting limitations in the model’s description of peripheral collisions.
A comparison with recent measurements at the same energy by the CMS Collaboration shows compatibility in both $R$ and $\lambda$ within 1.3$\sigma$. These results extend previous
\ensuremath{{\rm K}^{0}_{\rm{S}}}\xspace--\ensuremath{{\rm K}^{0}_{\rm{S}}}\xspace femtoscopy to a higher energy, providing a consistent baseline
for future comparisons.

\end{abstract}
\end{titlepage}

\setcounter{page}{2} %please do not remove this line

%%%%%%%%%%%%%%%%%%%%%%%%%%%%%%%%
% begin main text
%%%%%%%%%%%%%%%%%%%%%%%%%%%%%%%%

\section{Introduction} 
\label{sec:Introduction}
Understanding the space-time geometry of the collision region in high-energy heavy-ion collisions is essential for studying the properties of the particle-emitting source and subsequent particle interactions~\cite{intro1, intro2, intro3}. One of the primary techniques used to probe this source is femtoscopy, which examines momentum correlations between particles produced in such collisions~\cite{IdenticalBoson,Heinz}. By analyzing two-particle correlations at small relative momenta, femtoscopy provides insights into the size and evolution of the emission region. The shape of these correlations is influenced by both the source itself and the final-state interactions (FSI) between the particles, making femtoscopy a powerful tool for extracting information about the colliding system~\cite{IdenticalBoson}.

Identical boson femtoscopy has been used over the years to study the space-time geometry of the 
collision region of heavy-ion collisions because quantum statistical effects enhance correlation signals, making source size measurements more precise. Pions are the most commonly used particles due to their high production rates, allowing for detailed statistical studies~\cite{IdenticalBoson, PionExample}.
Identical kaon femtoscopy has been used to carry out studies of source size, final-state interactions, and hydrodynamic signatures such as
the measurements in \AuAu at \snn~$= 200$~\GeV by the STAR Collaboration~\cite{STARK0K0},
\pp collisions at \s~$= 5.02,~7$, and 13~\TeV, and \PbPb collisions at \twosevensixnn
by the ALICE Collaboration~\cite{ALICEK0K07, ALICEKchKch7, PRC276, PLB833}.

Neutral kaons exhibit several advantages for femtoscopic studies. 
They can be studied at higher 
momenta than charged kaons because neutral kaons can be 
identified using topological features of the decay geometry, 
that can be used reliably even at high momenta. Accessing the high-momentum region allows femtoscopic studies to probe effects of collective expansion and source dynamics over a broader kinematic range.
Additionally, neutral-kaon femtoscopy plays a key role in complementing charged-kaon femtoscopy studies.
Different identification and analysis methods are used for the two systems,
such as charged-particle tracking versus secondary-vertex reconstruction. There
is also a difference in methodology with the treatment of final-state interactions, since charged kaons incorporate the Coulomb force on top of strong interactions.

Femtoscopic studies of neutral-kaon pairs have established features
of collective radial flow and the effect of long-lived resonances on $\lambda$~\cite{STARK0K0, PRC276}. While these studies were carried out at
\snn = 200~\GeV and at \twosevensixnn, the beam-energy dependence of $R$ and $\lambda$
for neutral-kaon pairs at LHC energies remains to be addressed.

The goal of this letter is to extend previous \kzero--\kzero femtoscopy studies 
into a higher energy and analyze the dependence of the source size $R$ and the correlation strength $\lambda$ on the collision 
centrality and transverse momentum in \PbPb collisions at \fivenn. 
This study enables the investigation of the beam-energy dependence of $R$ and $\lambda$ by comparing them with the results from ALICE at \twosevensixnn~\cite{PRC276}.
A recent paper by the CMS Collaboration measured \kzero--\kzero, $\Lambda$--\kzero, and $\Lambda$--$\Lambda$ correlations in \PbPb collisions at \fivenn. For the \kzero--\kzero
results relevant to this work, a decreasing source size from about 4.6 fm
in central collisions to 1.6 fm in peripheral collisions is reported~\cite{CMS}.
A comparison of the new ALICE measurement to the CMS data will be provided later in this letter.
The present work also
includes a comparison to a hydrokinetic model with multiple stages: a high-density medium expansion described by ideal hydrodynamics, a gradual system 
decoupling with the hydrokinetic approach, and finally, a hadronic cascade within UrQMD~\cite{HKM, HKM502, URQMD}. The hydrokinetic model (HKM) was chosen because it incorporates both the hydrodynamics of the early stage of the collision and particle freeze-out, which
directly influence $R$ and $\lambda$. The integrated hydrokinetic model (iHKM)
extends the original HKM framework by incorporating pre-thermalization dynamics~\cite{HKM502}. Both models belong to a broader class of hybrid approaches that optimally describe each stage of the collision evolution by applying
the most appropriate theoretical framework at each phase, making them well-suited
for femtoscopic studies.

The organization of this letter is as follows:
in Sec.~\ref{sec:Data Analysis} the event and track selection criteria are described. In Sec.~\ref{sec:Measurment of CF} the theoretical and experimental
details of the correlation function and fitting procedure are discussed. Section~\ref{sec:Systematics} explains the various sources
and the procedure for calculating the systematic uncertainties. Finally,
the results are shown in Sec.~\ref{sec:Results} and summarized in Sec.~\ref{sec:Summary}.

\section{Data analysis}
\label{sec:Data Analysis}
This analysis uses 1.4 billion minimum bias events from \PbPb collisions at \fivenn measured with
the ALICE detector during the Run 2 data-taking period of the LHC.
The details and performance of the ALICE detector can be found in~\cite{ALICE2008, ALICEperformance}.
The minimum-bias trigger requires a signal in one of the forward scintillators that make up the V0 system. These detectors cover the pseudorapidity range
$-3.7 < \eta < -1.6$ and $2.8 < \eta < 5.1$~\cite{V0} and 
provide fast signals to trigger minimum-bias events.
The V0 is used to define centrality classes based on the amplitude it measures, which is proportional to charged-particle multiplicity. 
The detector response is studied using a sample of 6.7 million simulated events obtained using Monte Carlo (MC) techniques. The \PbPb collisions are generated with HIJING~\cite{HIJING} and the propagation of 
the particles through the ALICE detector is performed with GEANT4~\cite{GEANT}. 

The event- and track-selection criteria used in this analysis are identical to the ones given in Ref.~\cite{PRC276}, with changed parameters
discussed in the following.
The \kzero is not measured directly, instead, it is
reconstructed using its decay into charged pions.
The proper decay length for \kzero is $c\tau = 2.68$~cm and the branching ratio
for $\kzero\rightarrow\pip\pim$ is $(69.20 \pm 0.05) \%$~\cite{PDG}.
The Inner Tracking System (\ITS)~\cite{ALICE2008} is employed to determine the primary collision vertex (PV).
The $z$-position of the primary vertex, defined along the beam axis, is required to be
within $\pm 10$ cm of the nominal interaction point in the ALICE detector for the event to be included in the analysis. During track reconstruction, the PV is used as a spatial constraint
in the track-fitting procedure, improving the determination of track parameters.
The Time Projection Chamber (\TPC)~\cite{TPC} and the \ITS are used for the tracking of charged particles, allowing for
the reconstruction of the \kzero.
These two detectors cover the pseudorapidity range of \etarange{0.9}.

Both the \TPC and Time-Of-Flight (\TOF)~\cite{TOF}
detectors are used in the particle identification (PID) process. In the \TPC, the specific ionization
energy loss \dEdx is measured and used as the PID signal~\cite{ALICEperformance}.
The expected value of \dEdx for a given mass and momentum is calculated using a
parameterized Bethe-Bloch formula. For each particle species, the distribution
of the measured PID signal around the calculated value is fit with a Gaussian
to determine a width $\sigma$.
In the \TOF, the arrival time of the particle is measured in the 
pseudorapidity range \etarange{0.9}~\cite{TOF}. The particle time-of-flight is measured using the PV.
Similar to the \TPC, an expected time-of-flight is calculated for a given mass,
track length, and momentum. The spread of the measurement around the calculated
value is denoted by the Gaussian width $\sigma$.

For each track, the deviation in the number of $\sigma$s, $N_\sigma$, determining
how far away 
it is from the expected PID signal, is computed. For pion identification, the track is required 
to have a value of $N_\sigma$ less than 3 in the \TPC. For
tracks with momentum greater than 0.8~\GeVc, the \TOF is used in combination
with the \TPC, requiring an $N_\sigma$ lower than 3 also in the \TOF. At intermediate transverse-momentum
(\pt), up to 3 \GeVc, the \TPC and
\TOF provide more than $3\sigma$ separation power between pions, kaons, and protons,
ensuring effective PID~\cite{ALICEperformance}.

Along with the PID
requirements, the pion candidates are also required to have a \pt greater than 0.15~\GeVc to avoid tracking inefficiencies and material contamination at low \pt, \etarange{0.8} to avoid detector edge effects, and a three-dimensional (3D) distance of closest approach (DCA) to the PV greater than 0.4 cm to ensure the daughter tracks are not primary. Additionally, the DCA between the two daughter pions is also
required to be less than 0.3 cm to ensure a common decay vertex consistent with a \kzero decay. Finally, the pions are required to be oppositely charged to be able to form a \kzero.

After applying constraints to the pion candidates, the \kzero candidates are
subjected to their own selection criteria.
For this analysis, the \kzero candidates are restricted to a pseudorapidity range \etarange{0.8} 
to account for detector acceptance effects. A 3D DCA to the PV must be less than 0.3 
cm to ensure the \kzero is a primary particle. The decay length in the laboratory frame is 
required to be less than 30 cm. A longer decay length would recover more candidate \kzero, but
since tracking efficiency and momentum resolution worsen at large radii, the reconstruction
purity would degrade.
Finally, the 
reconstructed invariant mass is required to be within 0.485--0.510~\GeVmass.
If two reconstructed \kzero candidates share a daughter track, both are removed from the analysis. Only the events in which at least one \kzero--\kzero pair has been identified
are selected for the analysis.

\begin{figure}[!htbp]
    \centering
    \includegraphics[width=0.55\linewidth]{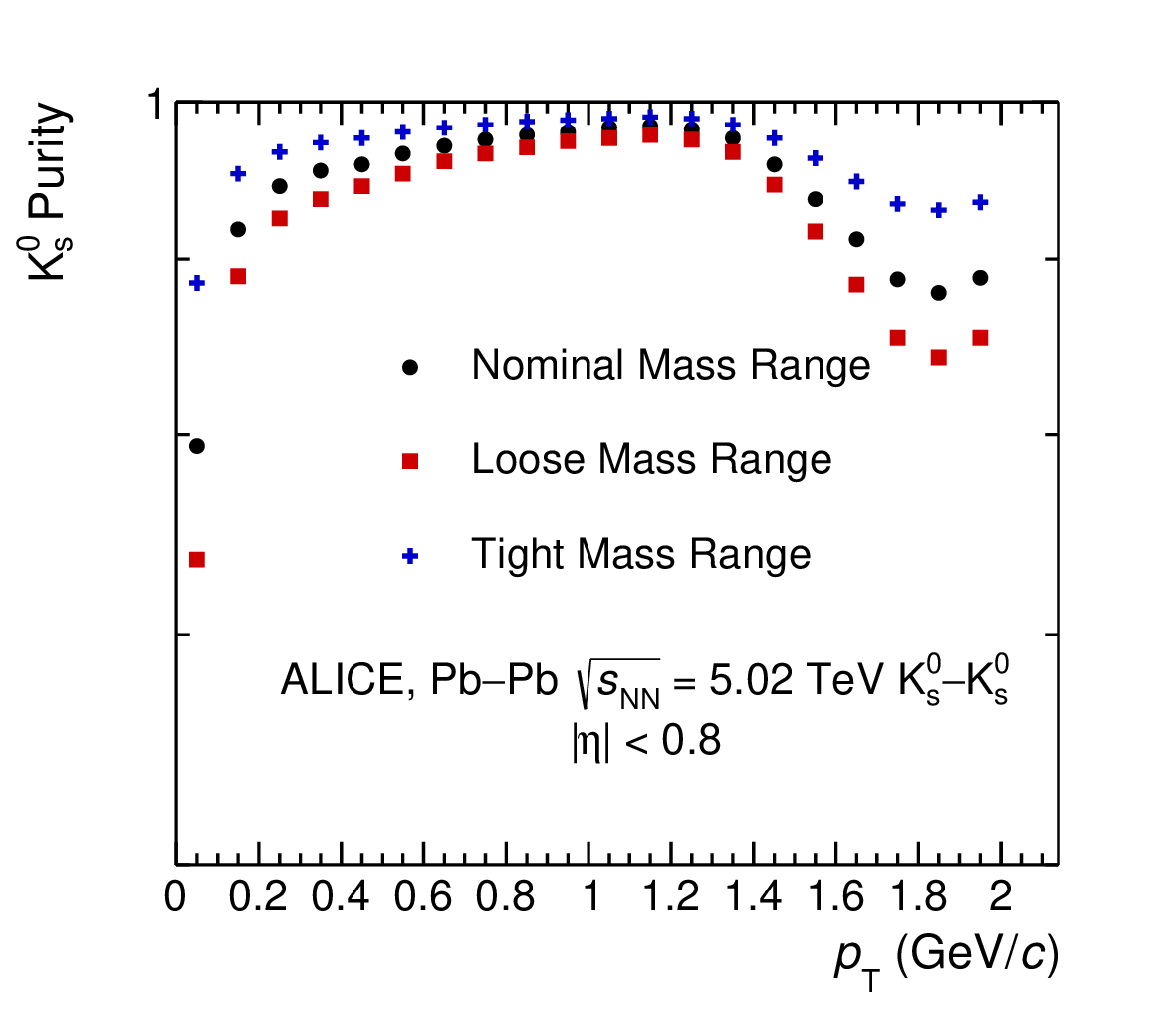}
    \caption{Purity of \kzero candidates as a function of \pt. Black markers represent the selection using the nominal invariant mass range $0.485 < M_{\pi\pi} < 0.510$~\GeVmass, red markers represent the selection using the loose range $0.480 < M_{\pi\pi} < 0.515$~\GeVmass, and blue markers represent the selection using the tight range $0.490 < M_{\pi\pi} < 0.505$~\GeVmass.}
    \label{fig:purity_pt}
\end{figure}

The topological requirements (\pip\pim DCA, $\pi$-vertex DCA, \kzero DCA, and decay length) used in this analysis are the same as those applied in the analysis of collisions at \twosevensixnn~\cite{PRC276} to ensure
consistency and a meaningful comparison between energies. In that analysis, the selection criteria were chosen based on ALICE MC simulations to balance signal purity and statistical yield. 
The purity of the reconstructed \kzero is defined as the ratio
signal/(signal + background). 
The invariant-mass distribution was integrated over the mass range to obtain the total signal + background.
The background is modeled by fitting a fourth-order polynomial to the 
side bands of the invariant-mass distribution (from 0.45--0.47 and 0.53--0.55~\GeVmass) outside the nominal
invariant-mass window (0.49--0.51~\GeVmass). 
The fitted polynomial is extrapolated into the signal region to provide a continuous
background across the full mass range. 
The signal is calculated by subtracting the background from the invariant-mass
distribution and then integrating over the signal window.
Using these values, the \pt-averaged purity was calculated to be 96\%.
The purity is not expected to change as a function of centrality,
as shown in~\cite{PRC276}.
Figure~\ref{fig:purity_pt} shows the purity as a function
of \pt for the three different invariant mass ranges considered in the systematic
uncertainty calculation.

In case of femtoscopic analyses involving primary tracks, two-track effects such as track merging (where two real tracks are reconstructed as one) and track splitting (where one real track is reconstructed as two) can distort the measured correlation function. Two-track effects are less prominent in \kzero--\kzero correlations
due to their indirect measurement, but still possible through close proximity of the daughter tracks. 
For each \kzero pair, the average distance between the same-sign daughter tracks 
is calculated by estimating the distance between the tracks at nine different 
radii throughout the \TPC volume (every 20 cm from 85 cm to 245 cm) and then averaging it.
The average separation of same-sign daughter tracks from the same event forming pairs is
divided by the same distribution of pairs coming from different events to minimize the 
combinatorial shape and extract the genuine two-particle effect. A depletion of this
ratio was observed for this analysis at separations of up to 6 cm.
To minimize this effect, each \kzero pair's same-sign daughters are required to have an average \TPC separation of at least 10 cm.

\section{Measurement of the correlation function}
\label{sec:Measurment of CF}
The experimental correlation function is defined
as $C(k^*) = NA(k^*) / B(k^*)$, where $A(k^*)$ is
the measured relative momentum distribution of pairs from the same event which could be correlated, $B(k^*)$ is the measured momentum distribution of pairs from different events (one \kzero
comes from one event and is paired with another \kzero coming from a different event)
which are not correlated, and $N$ is a normalization constant to account for
differences in the total number of pairs between $A(k^*)$ and $B(k^*)$. Event
mixing was performed using pools containing 100 events.
Here, $k^*$ is the momentum of one of the particles in the pair rest frame, defined as the frame where $p_1^* = p^*_2 = k^* = |\vec{k^*|}$~\cite{PDG}, where the
asterisk denotes quantities evaluated in this frame. Pairs made from the same
event contain the correlation signal arising from particle interactions, 
while mixed-event pairs account only for the phase-space. Only the events which are similar in 
centrality percentage and position of primary vertex 
are combined in the event-mixing and this is ensured
by pairing two \kzero only if the difference in $z$ positions of the vertices and centrality percentages of the corresponding events
are within 2 cm and 2\% of each other, respectively. The requirement for same-sign daughters to have a 
minimum average separation of 10 cm in the \TPC is also applied for mixed events.

For the identical kaon case, the masses of the two particles are the same, leading to $k^* = q_{\rm inv}/2$. The square of the invariant momentum difference is given by $q_{\rm inv}^2 = |\vec{p_1} - \vec{p_2}|^2 - |E_1 - E_2|^2$, evaluated with momenta and energies measured in the laboratory frame.

This analysis is carried out in several centrality and \kt ($\kt = |\vec{p}_{\rm T,1} + \vec{p}_{\rm T,2} | /2$) ranges. The centrality and \kt ranges are kept the same as those of the previous analysis
at \twosevensixnn~\cite{PRC276} so that the results can be directly compared.
The four centrality and four \kt 
ranges are: 0--10\%, 10--30\%, 30--50\%, and 50--90\%, and $0.2 < \kt < 0.6, 0.6 < \kt < 0.8, 0.8 < \kt < 1.0$, and $1.0 < \kt < 1.5$~\GeVc, respectively.

For this analysis, the data collected between 2015 and 2018 are combined. The individual same-event pair distributions, mixed-event pair distributions, and correlation functions obtained from each dataset are compatible in shape within statistical uncertainties, which justifies combining them to increase the statistical significance. The combination is performed by first summing the same-event pair distributions from the datasets and then summing the mixed-event pair distributions. A final correlation function is constructed by dividing these summed numerators and denominators.

Finite track resolution arises from the limited precision of the tracking
detectors, which leads to a smearing of the momentum distribution of the
reconstructed tracks. This resolution effect propagates into the calculation of 
the relative momentum between particle pairs, possibly distorting the measured
correlation function.
To account for the effect of smearing, MC simulations are employed.
In these simulations, two versions of each particle's momentum are available:
the ``true'' momentum ($k^*_{\rm true}$) taken directly from the HIJING
generator-level output and the ``reconstructed'' momentum ($k^*_{\rm reco}$)
obtained after simulating the detector response and applying the full 
reconstruction chain used for data, including track fitting and PID.
Two correlation functions are constructed using the true and reconstructed momenta separately. Since HIJING does not incorporate
quantum statistics or final-state interactions effects, the same-event distributions
are weighted with a Gaussian form: $1 + \lambda e^{-4R^2{k^*}^2}$, assuming specific input values for $R$ and $\lambda$ chosen to be consistent with the \twosevensixnn results. 
The Gaussian form was chosen because it is the same as the quantum 
statistics term, which provides the leading 
contribution to the correlation signal~\cite{QS1, QS2}. The mixed-event distributions 
are filled with unit weights. The ratio of the ``true'' correlation function 
to the ``reconstructed'' one is applied to the data as a bin-by-bin
correction factor~\cite{PLB833}. The correction factor is largest at low $k^*$ and has 
a negligible effect at higher values. The overall correction is below $2\%$ difference on the extracted values of $R$ and $\lambda$ from the fit procedure described in the next section.

\subsection{Modeling the correlation function}

The \kzero--\kzero correlation functions are fitted with a model that accounts for both the quantum statistics and strong interaction effects. This study adopts the Lednický-Lyuboshitz (LL) analytical formula,
which assumes a Gaussian shape of the source~\cite{Lednicky1, Lednicky2, Lednicky3}.
For this study, the non-femtoscopic background is described by a first-order
polynomial with parameters $b$ and $c$. This background consists of correlations
that do not arise from FSI effects and are often caused by momentum conservation effects~\cite{PRC276}. The
full function used to fit the experimental correlation function reads

\begin{equation}
    C_{\text{fit}}(k^*) = N (b k^* + c) C_{\text{LL}}(k^*).
    \label{eq:Cfit}
\end{equation}

The normalization factor $N$ accounts for overall scaling, while baseline
parameters $b$ and $c$ account for linear background behavior. The general form of the LL correlation function is
\begin{equation}
\begin{split}
    C_{\text{LL}}(k^*) = 1 + \lambda e^{-4k^{*2}R^2} + \lambda \alpha \left[ \left|\frac{f(k^*)}{R}\right|^2 + \frac{4\Re f(k^*)}{\sqrt{\pi}R}F_1(2Rk^*) - \frac{2\Im f(k^*)}{R}F_2(2Rk^*)  \right].
\end{split}
\label{eq:Led}
\end{equation}

The first term in the equation, $1 + \lambda e^{-4k^{*2}R^2}$, represents the
contribution arising from quantum statistics, which is needed because the \kzero--\kzero system consists of identical bosons. In the ideal case without long-lived
resonances or experimental effects, $\lambda$ is expected to equal 1. This is true
even with underlying {\ensuremath{{\rm K}^{0}}\xspace}--{\xspace\ensuremath{{\rm\overline{K}}^{0}}\xspace} mixing, as the reconstructed
\kzero--\kzero pair is an identical-boson system~\cite{Lednicky3}.
The remaining terms make up the 
FSI contribution term, in which $\alpha$ takes into account
the {\ensuremath{{\rm K}^{0}}\xspace}--{\xspace\ensuremath{{\rm\overline{K}}^{0}}\xspace} abundance asymmetry and is set to 0.5 assuming symmetry in {\ensuremath{{\rm K}^{0}}\xspace}--{\xspace\ensuremath{{\rm\overline{K}}^{0}}\xspace}
production~\cite{symmetry}. 
The $F_1$ and $F_2$ terms arise from the integration of the two-particle wavefunction
over a Gaussian source and have the following forms
\begin{equation}
    F_1(z) = \int_0^z dx \frac{e^{x^2 - z^2}}{z}
    \label{eq:F1}
\end{equation}
and
\begin{equation}
    F_2(z) = \frac{1 - e^{-z^2}}{z}.
    \label{eq:F2}
\end{equation}

The \kzero is represented by the eigenstate 
$\ket{\kzero} = \frac{1}{\sqrt{2}}({\ensuremath{{\rm K}^{0}}\xspace} + {\xspace\ensuremath{{\rm\overline{K}}^{0}}\xspace})$,
implying the two-particle \kzero--\kzero system can be decomposed into
total isospin $I = 0$ and $I = 1$ channels. Both channels are accounted for in the total scattering amplitude

\begin{equation}
    f(k^*) = \frac{f_0(k^*) + f_1(k^*)}{2}.
    \label{eq:f(k)}
\end{equation}

The s-wave scattering amplitude $f(k^*)$ is dominated by the isoscalar $f_0$ and 
isovector $a_0$ resonances
and is written in terms of the resonance masses and decay couplings~\cite{Lednicky3}.
The $f_0$ resonance has a mass of 990 $\pm$ 20~\MeVmass and the $a_0$ has a mass of 980 $\pm$ 20~\MeVmass,
suggesting that they both lie slightly below the \kzero--\kzero threshold~\cite{PDG}.
The non-resonant FSI is expected to be negligible in comparison
to the $a_0$ and $f_0$ contributions~\cite{STARK0K0, resonance}.
The individual amplitudes are written as

\begin{equation}
    f_I(k^*) = \frac{\gamma_I}{m_I^2 - s - i(\gamma_I k^* + \gamma_I' k_I')}.
    \label{eq:fI(k)}
\end{equation}

Equation~\ref{eq:fI(k)} is computed for two values of $I$, $I = 0$ for the
$f_0(980)$ resonance and $I=1$ for the
$a_0(980)$ resonance. Here, $m_I$ is the mass of the corresponding resonance and $\gamma_I$ and $\gamma_I'$ 
are the couplings of the resonance to the kaon or other channels. These coupling constants are related to the resonance’s partial decay widths, and hence to their branching ratios into each channel. For the $a_0$ and $f_0$ resonances, the other channels are $\pi\eta$ and $\pi\pi$, respectively. Additionally,
$s = 4(m_K^2 + k^{*2})$ and $k_I'$ denotes the momentum in the secondary decay channel.
Although many different resonance masses, couplings, and momenta have been used in other
analyses, the ``Achasov2'' parameters are used in this study, which have been shown to be optimal~\cite{PRC107, PLB774}. These parameters, all in \GeV, are: $m_{f_0} = 0.967$, $\gamma_0 = \gamma_{f_0 \rm K \overline{\rm K}} = 0.34$, $\gamma_0' = \gamma_{f_0 \pi \pi}' = 0.089$, $m_{a_0} = 1.003$, $\gamma_1 = \gamma_{a_0 \rm K \overline{\rm K}} = 0.8365$, and 
$\gamma_1' = \gamma_{a_0 \pi \eta}' = 0.4580$~\cite{Achasov2}.

A fit to the measured correlation function is performed in order to obtain the source
size $R$ and the so-called correlation strength $\lambda$.
The parameter $\lambda$ quantifies the fraction of particle pairs that contribute
to the genuine femtoscopic correlation and is reduced from unity by 
experimental effects and non-primordial particle production.
It is defined as $\lambda = P_i f_i^{\rm prim}  P_j f_j^{\rm prim}$~\cite{intro2, lambda2}, where
$P_{i,j}$ are the purities of the selected \kzero and $f_{i,j}^{\rm prim}$
are the fractions of \kzero directly produced
in the initial collision leading to the correlation signal.
The primordial fraction takes into account the dilution of the correlation due to strongly decaying
resonances such as the $\phi(1020)$ and $\rm K^{*}(892)$~\cite{PLB833}.
More details on the dilution effect from these resonances will be discussed in Sec.~\ref{sec:Results}. The parameters $N$, $b$, and $c$ in Eq.~\ref{eq:Cfit} are also
obtained from the fit simultaneously to the measurement of $R$ and $\lambda$.

\begin{figure}[!htbp]
    \centering
    \includegraphics[width=0.9\linewidth]{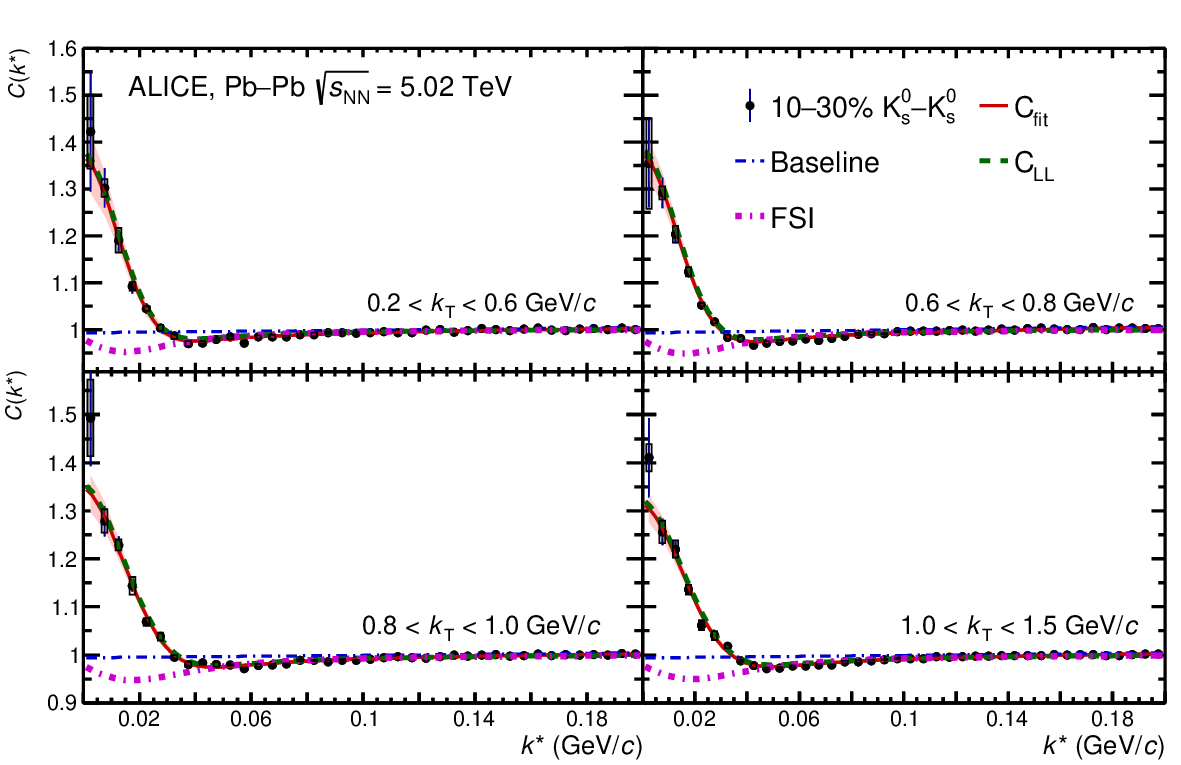}
    \caption{Correlation functions with the respective fits in four \kt ranges for \PbPb collisions at \fivenn in the 10-30\% centrality range. Statistical uncertainties are shown as the black bars and systematic uncertainties, described in Sec.~\ref{sec:Systematics}, are shown as boxes. The linear baseline contribution is shown in blue and the Lednický-Lyuboshitz model from Eq.~\ref{eq:Led} is shown in green. The total fit, shown in red, corresponds to Eq.~\ref{eq:Cfit} and is obtained by fitting the product of the baseline and the Lednický-Lyuboshitz model to the data. The shaded band around the fit
    represents the 95\% confidence interval, reflecting statistical uncertainty
    on the fit parameters.}
    \label{fig:CFs}
\end{figure}

Figure~\ref{fig:CFs} shows a representative example of measured correlation functions
and the corresponding fit using Eq.~\ref{eq:Cfit}. As illustrated, the chosen model
provides a good description of the data across the full $k^*$ range, with 
a typical $\chi^2/\rm{ndf} = 1.22$ in the range of $k^*$ from 0 to 0.2~\GeVc.
Correlation functions and fits for the other centrality ranges can be found in Appendix~\ref{app:fits}, along with a summary table of all fit parameters in Table~\ref{tab:LSys}.

\section{Systematic uncertainties}
\label{sec:Systematics}
The systematic uncertainties on the measured correlation function were evaluated
as a function of $k^*$. Only the variations that directly affected the
value of $C(k^*)$ were included in this procedure. For each selection criterion,
the correlation functions were recomputed and the value of $C(k^*)$ was compared to the nominal result.
For a given value of $k^*$, the systematic uncertainty was quantified as the RMS of
the deviations obtained from all relevant variations. These uncertainties can be
seen in Fig.~\ref{fig:CFs} and the other figures in Appendix~\ref{app:fits} as the empty boxes
around the data markers.

The momentum resolution procedure described in Sec.~\ref{sec:Measurment of CF} was repeated for different input values of $R$ and $\lambda$, and the resulting variation was found to be negligible.

The sources of systematic uncertainties on $R$ and $\lambda$
consider the selection criteria
such as the track merging requirement, the difference in centrality for mixing events, and the choice 
of the invariant mass window of the \kzero mesons. 
Variations in the DCA to the PV of the \kzero and the $N_\sigma$ selection for pion PID
in the TPC are also applied.
Additionally, the range of fitting the correlation function in $k^*$ is also shown as a contributing factor to the systematic uncertainty. 
For the PID and centrality matching tolerance of the mixed events variations, only a tightening of the values
was considered. Loosening these selections would result in a lower purity or less
accurate event pairing and will not be considered.
A summary of the selection criteria can be found in Table~\ref{tab:SysParams}.

\begin{table}[!htbp]
 \centering
  \caption{Selection and fitting criteria varied to determine the systematic uncertainty on \R and $\lambda$.}
    \begin{tabular}{ cccc } 
     \hline
     Systematic variation & tight & nominal & loose \\
     \hline
     $k^*$(\GeVc) fit range & 0.08 & 0.10 & 0.12\\
     
     Track merging selection (cm) & 8 & 10 & 12\\

     Event-mixing centrality matching tolerance (\%) & 1 & 2 & -\\
     
     \kzero mass (\GeVmass) & 0.490--0.505 & 0.485--0.510 & 0.480--0.515\\
     
     \kzero DCA to PV (cm) & 0.25 & 0.30 & 0.35\\
     
     TPC $N_\sigma$ for pions & 2.5 & 3.0 & -\\
     \hline
    \end{tabular}

 \label{tab:SysParams}
\end{table}

Individual contributions to the systematic uncertainty were calculated by varying one
selection criterion at a time, while keeping all others fixed. 
In each of the 16
centrality--\kt ranges, new fits to the correlation functions were performed and 
the deviations from the nominal $R$ and $\lambda$ values were calculated. For a given
selection criterion, the change in extracted parameters is quantified by the
root-mean-square (RMS) of these deviations, calculated from the deviations of the loose and tight variations relative to the nominal value when both are available, or from the deviations of the nominal and varied values with respect to their average when only a single variation is performed.
The individual sources of uncertainty for each centrality and \kt range are treated
as uncorrelated and the RMS values are added in quadrature to obtain the total
systematic uncertainty.
As seen throughout many of the individual
systematic checks, the total uncertainty grows for more peripheral cases. 
The systematic uncertainties are in some cases larger than those reported in the 
2.76~\TeV study~\cite{PRC276} as the additional sources of uncertainty
(pion PID and background mixing requirements) were not considered in the previous study, leading
to a possible underestimation of systematic uncertainty.
These sources were included here because they provide significant contributions to the 
overall uncertainty.
For $R$, the variation with the largest contribution is the DCA to the PV of the \kzero, which introduces an uncertainty of 13.6\% for the 50--90\% and $1.0 < \kt < 1.5$~\GeVc fit. 
For $\lambda$, the dominant source of uncertainty is the track-merging effect selection criteria, which
contributes with 0.35\% to 28.2\% depending on centrality and \kt range.

\section{Results and discussion}
\label{sec:Results}

\begin{figure}[!htbp]
    \centering
    \includegraphics[width=1.0\linewidth]{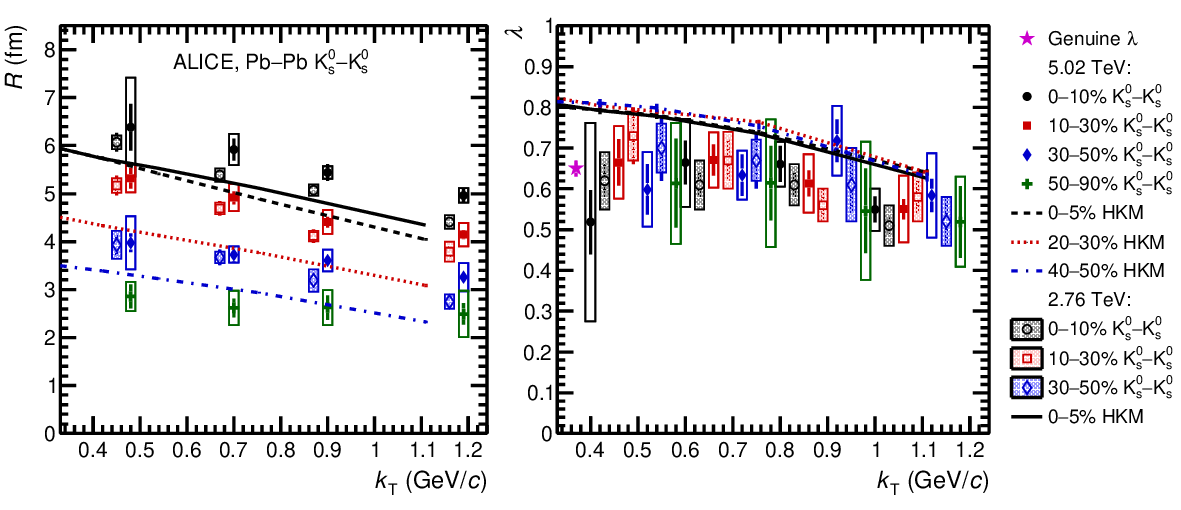}
    \caption{Extracted radii $R$ (left panel) and $\lambda$ (right panel) as a function of the average pair-transverse momentum \kt in the ranges 0.2--0.6, 0.6--0.8, 0.8--1.0, and 1.0--1.5~\GeVc for different centrality classes. Filled markers represent the results obtained for \PbPb at \fivenn, hollow markers refer to results obtained in \PbPb at \twosevensixnn~\cite{PRC276} with a horizontal offset applied for legibility. Vertical bars and boxes represent the statistical and systematic uncertainties, respectively. Predictions from HKM model calculations are reported with lines for different center-of-mass energies and centrality ranges. In the right panel, the star marker represents the data-driven calculation of the genuine $\lambda$ value (evaluated independently of \kt) from resonance-decay contributions. See text for details.}
    \label{fig:RkT}
\end{figure}

The left side of Fig.~\ref{fig:RkT} shows the extracted radii as a function of
average pair \kt in the given range
for different centrality classes.
As collisions become more peripheral and there is less overlap
between the incoming nuclei, the radius of the source decreases.
The results for the radii at \fivenn and 2.76~\TeV in these three
multiplicity classes are in agreement within a maximum deviation of 1.6$\sigma$, 
occurring in the 0--10\% centrality range.
All quoted deviation values are obtained by combining the deviations of multiple
data points within a given centrality interval.
This indicates that the space-time structure and
expansion do not change drastically as a function of collision energy at high multiplicity.
For the first three
centrality classes (0--10\%, 10--30\%, and 30--50\%)
a decrease in radii is also seen with increasing \kt, which qualitatively agrees with the
\twosevensixnn results. This decrease with \kt was quantified by fitting the $R(\kt)$
values with both a constant and a linear function for each centrality range.
In the first three centrality
ranges, the linear fit yields significantly smaller $\chi^2/\rm{ndf}$ values than the constant fit, consistent with the decrease of
measured radii with increasing \kt as seen in Table~\ref{tab:lin vs flat}.
The HKM predictions show a clear decrease of $R$ with increasing \kt across all the
centrality ranges, further supporting the linear description over a constant one.

\begin{table}[!htbp]
    \centering
    \caption{$\chi^2/\rm{ndf}$ values for constant and linear fits of $R$ as a function of \kt for all four centrality ranges.}
    \begin{tabular}{ccc}
        \hline
        Centrality & Constant $\chi^2/\rm{ndf}$ & Linear $\chi^2/\rm{ndf}$ \\
        \hline
        0--10\% & 6.0 & 0.09 \\
        10--30\% & 13.0 & 1.1 \\
        30--50\% & 4.2 & 0.09 \\
        50--90\% & 0.41 & 0.12 \\
        \hline
    \end{tabular}
    \label{tab:lin vs flat}
\end{table}

This flow-driven reduction in radii at higher \kt 
has been widely seen across many femtoscopic studies for $\pi^\pm\pi^\pm$, $\rm K^\pm \rm K ^\pm $, \pp, and $\pbar\pbar$ correlations~\cite{PRC276}, confirming that the 
\kzero source behaves similarly to other hadronic species.
However, for more peripheral collisions (50--90\% centrality), the slope of $R$ as a function of \kt appears flatter than in central collisions, indicating possibly weaker radial flow at lower
multiplicities. 
It has been shown that at large $\beta_T = 2\kt / (E_1 + E_2)$, $q_{\rm inv}$ and similarly $k^*$ is not an optimal 1D variable because it does not fully capture the 3D
structure of the source~\cite{1Dqinv}. This effect is expected to be more pronounced
at higher \kt and may contribute to the apparent flattening of $R$ in
peripheral collisions.
This behavior mirrors a similar trend seen in \pp collisions and suggests that the system at lower multiplicity transitions toward more \pp-like behavior~\cite{ppBehavior}.
This trend is consistent with the expected development of radial flow
in the expanding medium, where more central collisions generate larger
systems with stronger flow~\cite{radial_flow}.

The left panel of Fig.~\ref{fig:RkT} also shows a comparison between the measured radii and two hydrokinetic model predictions at different collision energies: a 
prediction at 2.76~\TeV 
for the most central events (0–5\%), and predictions at 5.02~\TeV for three centrality 
classes (0–5\%, 20–30\%, and 40–50\%)~\cite{HKM, HKM502}, with statistical uncertainties included.
The HKM calculations are obtained by applying a \pt selection on \kzero of $0.14 < \pt < 1.5$~\GeVc,
while no explicit \kzero \pt selection is applied in this analysis.
A cross-check was performed by applying the HKM \pt selection to the data, and
the extracted $R$ and $\lambda$ values were found to be compatible with the
nominal results within uncertainties.
For the 0--10\% centrality range \PbPb collisions at both energies, the model 
reproduces the measured radii reasonably well. 
However, in the 30--50\% centrality range at 5.02~\TeV, the measured radii deviate from the model
predictions by up to 3.6$\sigma$ and appear to decrease less steeply
than the predictions with increasing \kt.
This suggests that the model may overestimate the strength of collective flow in low-multiplicity events.

The $\lambda$ parameter is also shown as a function of \kt
for the different energies and centralities, as seen in the right panel of Fig.~\ref{fig:RkT}. Here, the values are corrected by dividing by the
calculated purity from Sec.~\ref{sec:Data Analysis} in order to remove the dilution
effect from particle misidentification. The remaining reduction in the value of the extracted
$\lambda$ comes from long-lived resonances, primarily the $\rm K^{*}(892)$ and $\phi(1020)$,
which are discussed below.
The values of $\lambda$ at 5.02 and 2.76~\TeV agree within uncertainties across most
centrality intervals and \kt ranges. The HKM predictions for $\lambda$ tend to 
overestimate the measured values, particularly at lower \kt.
This points to a potential
oversimplification in the model's treatment of source purity, especially at low pair
momentum.
The model explicitly considers contributions from
resonances such as the $\rm K^{*}(892)$ and $\phi(1020)$, which decay into kaons. As seen in Ref.~\cite{HKM}, excluding
these resonances does not have a significant effect on the
extracted radii or $\lambda$ values.
The slightly negative slope predicted by the model at higher \kt is
compatible within uncertainties with the measured $\lambda$.

The dilution effect on the purity-corrected $\lambda$ parameter due to the strong
resonances $\rm K^{*}(892)$ and $\phi(1020)$~\cite{PDG} decaying into \kzero can
be estimated directly from ALICE measurements~\cite{PbPbratios}, following the
procedure adopted from \pp collisions~\cite{PLB833}. By taking the measured
ratios for both $\rm K^{*0}/\kzero$ and $\phi/\kzero$, the centrality-averaged 
primordial \kzero fraction is estimated to be $f_{\kzero}^{\rm prim} = 0.81 \pm 0.02$
(see Ref.~\cite{PLB833} for details). Hence, the corresponding centrality-averaged genuine $\lambda$
parameter is expected to be equal to 0.65 $\pm$ 0.02. As shown in Fig.~\ref{fig:RkT}
(right panel), this estimate is consistent, within uncertainties, with the extracted
$\lambda$ values in this work when assumed as free parameters. This indicates that the 
majority of the \kzero--\kzero pairs entering in the measured correlation functions
are primordially produced in the initial collision.

\begin{figure}[!htbp]
    \centering
    \includegraphics[width=1.0\linewidth]{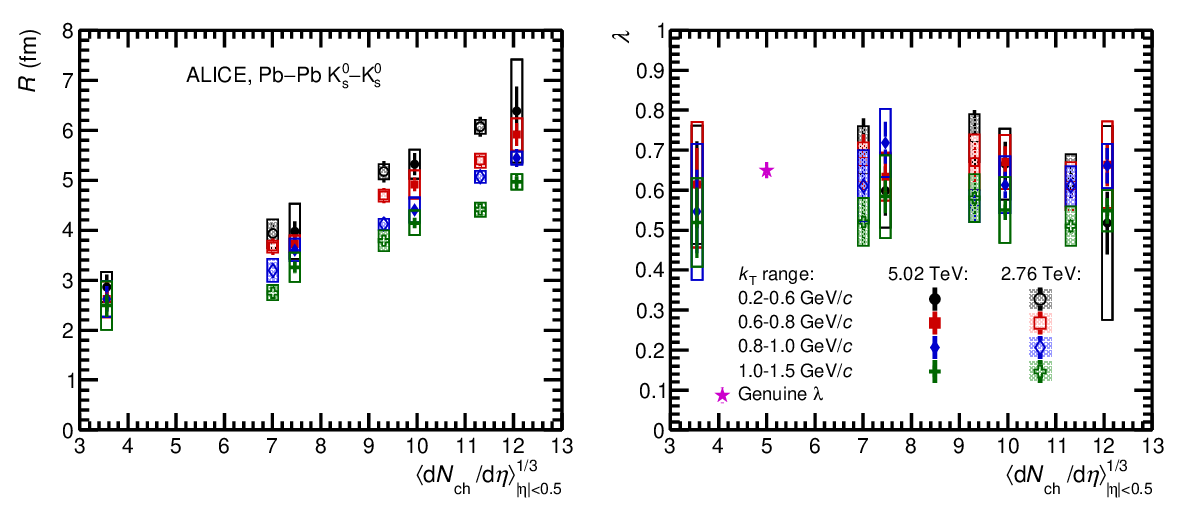}
    \caption{Radii $R$ (left) and $\lambda$ (right) as a function of the cubed root of the average charged-particle multiplicity density for different \kt ranges. Filled markers represent the results obtained for \PbPb at \fivenn, hollow markers refer to results obtained in \PbPb at \twosevensixnn~\cite{PRC276}. Vertical bars and boxes represent the statistical and systematic uncertainties, respectively. In the right plot, the star marker represents the estimated genuine $\lambda$ value calculated from resonance decay contributions. See text for details.}
    \label{fig:RMult1}
\end{figure}

This study focuses on comparing the results in the same colliding system but at different center-of-mass energies. To address the differences in multiplicity, the
results are shown as a function of the cubed root of the average charged-particle multiplicity density $\avdndeta ^{1/3}$~\cite{Mult276, Mult502}. 
This choice is motivated by the scaling of charged-particle multiplicity density with the source volume, meaning $\avdndeta ^{1/3}$ is expected to scale linearly with $R$~\cite{IdenticalBoson}. This assumes that the source is well described by a 
single Gaussian radius, with the 1D analysis representing well the 3D case.
The left panel of Fig.~\ref{fig:RMult1} shows an increase in radius
with an increase in the charged-particle multiplicity density for all \kt ranges
in both energies. This reflects the simple geometry of the emitting source, where a larger
initial overlap of the collision results in a larger radius parameter $R$.
The measured radii at 5.02 and 2.76~\TeV agree well within uncertainties for the 
three most central classes, which is consistent with the idea that the initial geometry 
dominates the system size at high multiplicity. The maximum deviation between radii
extracted at the two different center-of-mass energies is of 1.6$\sigma$.

Comparisons between the values of $\lambda$ for the two different 
collision energies as a function of
the charged-particle multiplicity density are also presented in the right panel of Fig.~\ref{fig:RMult1}. 
The $\lambda$ values show consistency across the studied range, with no
significant trends observed. The results at both energies are in agreement
within uncertainties.
The maximum deviation between the values of $\lambda$ is 0.51$\sigma$. In general, the value of $\lambda$ stays stable
around 0.6 for all energies, \kt, and multiplicity density.
The stability of $\lambda$ suggests that the relative contributions of primordial
and non-primordial \kzero remains largely unchanged across collision energies.

\begin{figure}
    \centering
    \includegraphics[width=0.95\linewidth]{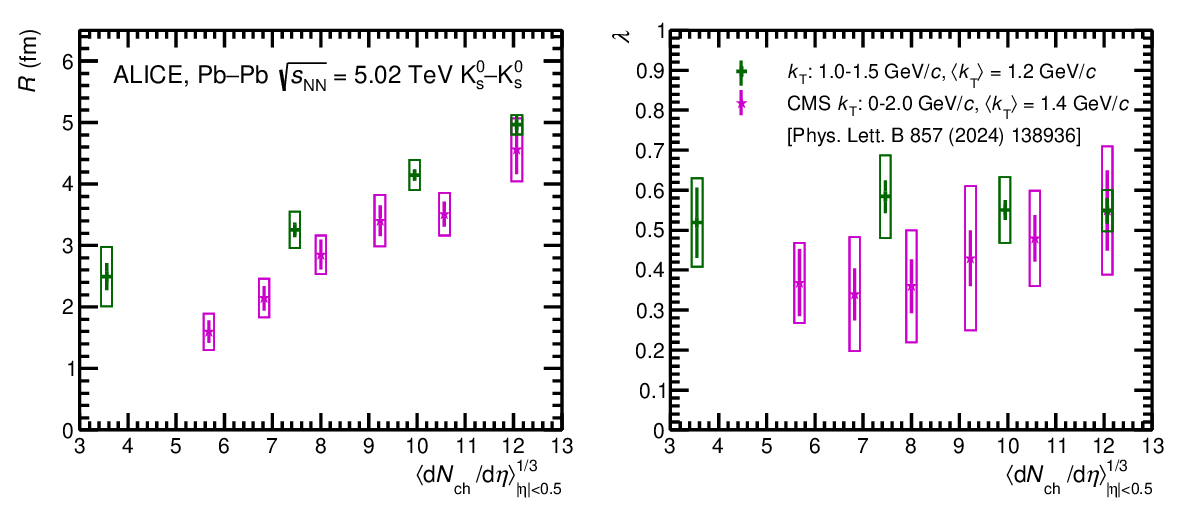}
    \caption{Radius (left) and $\lambda$ (right) parameters as a function of $\avdndeta^{1/3}$ from the highest \kt range compared with results from CMS.}
    \label{fig:CMScomps}
\end{figure}

A comparison is made between the results of this study and those recently published by the 
CMS Collaboration~\cite{CMS}, which analyzes \kzero--\kzero in the same collision system and energy, with the extracted source parameters shown as a function of
$\avdndeta^{1/3}$ in Fig.~\ref{fig:CMScomps}. The CMS Collaboration uses a broader transverse momentum range ($0 < \kt < 2.0$~\GeVc) than this study. 
To closely match the wide \kt range used by CMS, the comparison in this study is limited
to the highest \kt interval, corresponding to an average transverse mass of 
$\langle m_{\rm T} \rangle = 1.3$~\GeVmass, to be compared with the reported CMS value of
1.5~\GeVmass~\cite{CMS}. The two experiments also differ in their choices of 
baseline. This study uses a linear baseline as shown in Eq.~\ref{eq:Cfit} and CMS
uses a double Gaussian form~\cite{CMS}. The impact of this difference was tested
by refitting the correlation function using a double Gaussian baseline and
the effect on $R$ and $\lambda$ was found to be negligible.
The two experiments are in qualitative agreement, confirming overall consistency.
The results for $R$ and $\lambda$ obtained by ALICE in this study are consistent with
those measured by CMS, within an average $1.3\sigma$ deviation in the overlapping
multiplicity density range.
Within the multiplicity
range shared by both measurements, the present results have smaller total
uncertainties than those reported by CMS, indicating improved precision.

\section{Summary}
\label{sec:Summary}
This study presents \kzero--\kzero femtoscopy measurements in \PbPb collisions at \fivenn by the
ALICE Collaboration at the LHC.
The results extend the study of the emitting source size in 
Pb--Pb collisions to higher center-of-mass energies compared to previously published
results at \twosevensixnn~\cite{PRC276}.
The femtoscopic radii $R$ and correlation strengths 
$\lambda$ are extracted as a function of pair-transverse momentum (\kt) and 
charged-particle multiplicity density, providing insight into the space-time structure and dynamics 
of the particle-emitting source.

The radii exhibit a clear centrality and \kt dependence consistent with the expectations from 
collective flow: larger $R$ values are observed in central collisions, and radii decrease with increasing \kt in high-multiplicity events. This behavior is qualitatively and quantitatively 
consistent with previous ALICE results at 2.76~\TeV, indicating that the 
collective expansion reflected in the extracted $R$ values shows no significant energy
dependence at high multiplicity. However, in peripheral collisions, 
the dependence of $R$ as a function of \kt appears to flatten, possibly suggesting weaker radial flow and a possible 
transition toward pp-like behavior.

The purity-corrected $\lambda$ parameters remain largely consistent across \kt, centrality, and collision 
energy, suggesting that the underlying emitting source and correlation strength are 
relatively unchanged between 2.76 and 5.02~\TeV energies. 
The extracted values of $\lambda$ are overall in agreement with data-driven
estimates on the primordial fractions of \kzero--\kzero pairs produced in the collision.
This stability indicates that the 
correlation strength is not strongly affected by the collision energy within the measured phase space.

A comparison with published CMS results at the same center-of-mass energy shows 
compatibility with the results obtained in this work, with deviations not exceeding $1.3\sigma$
across the studied multiplicity range. This study complements and
extends the results obtained by the CMS Collaboration with improved precision.

Together, these results provide further evidence of collective behavior in high-energy heavy-ion collisions, with the \kt dependence of $R$ becoming less pronounced
in peripheral collisions. 
These findings improve our understanding of the space–time structure of the particle-emitting source and provide valuable constraints for theoretical models of heavy-ion collisions. Future studies with higher statistics and expanded kinematic coverage will help further elucidate the dynamics of the system in peripheral collisions and refine comparisons across experiments.

%%%%%%%%%%%%%%%%%%%%%%%%%%%%%%%%
% end main text 
%%%%%%%%%%%%%%%%%%%%%%%%%%%%%%%%

%%%%% acknowledgements - handled by EB chairs 
\newenvironment{acknowledgement}{\relax}{\relax}
\begin{acknowledgement}
\section*{Acknowledgements}
% add specific acknowledgements here 
% ...but please don't remove the line below: funding agencies
% will be acknowledged with a custom tex file handled by EB chairs after Collab Round 2
\input{fa_2026-03-26_Opt_C.tex}
\end{acknowledgement}

%%%%%%%% Bibliography 
\bibliographystyle{utphys}   % Remember we use title in the biblio
\bibliography{bibliography}
%\input {bibliography.tex}  

%%%%%%%%%%%%%%%%%%%%%%%%%%%%%%%%
% Appendices: yours (if any) + authorlist
%%%%%%%%%%%%%%%%%%%%%%%%%%%%%%%%
\newpage
\appendix

\input{CFappendix} % put your appendices here (if any)
\clearpage

%%%%% Authorlist - please do not touch: handled by EB chairs 
\section{The ALICE Collaboration}
\label{app:collab}
\input{Alice_Authorlist_2026-03-26_Opt_C.tex}
\end{document}

%% file: commands.tex
%%%%%%%%%%%%%%%%%%%%%%%%%%%%%%%%%%%%%%%%%%%%%%%%%%
% These are some new commands that may be useful 
% for paper writing in general. If other newcommands
% are needed for your specific paper, please feel 
% free to add here. 
%
% The currently available commands are organized in: 
% 1) Systems
% 2) Quantities
% 3) Energies and units
% 4) Detectors
% 5) particle species 
%%%%%%%%%%%%%%%%%%%%%%%%%%%%%%%%%%%%%%%%%%%%%%%%%%

% 1) SYSTEMS 
\newcommand{\pp}           {pp\xspace}
\newcommand{\ppbar}        {\mbox{$\mathrm {p\overline{p}}$}\xspace}
\newcommand{\XeXe}         {\mbox{Xe--Xe}\xspace}
\newcommand{\PbPb}         {\mbox{Pb--Pb}\xspace}
\newcommand{\pA}           {\mbox{pA}\xspace}
\newcommand{\pPb}          {\mbox{p--Pb}\xspace}
\newcommand{\AuAu}         {\mbox{Au--Au}\xspace}
\newcommand{\dAu}          {\mbox{d--Au}\xspace}

% 2) QUANTITIES 
\newcommand{\s}            {\ensuremath{\sqrt{s}}\xspace}
\newcommand{\snn}          {\ensuremath{\sqrt{s_{\mathrm{NN}}}}\xspace}
\newcommand{\pt}           {\ensuremath{p_{\rm T}}\xspace}
\newcommand{\kt}           {\ensuremath{k_{\rm T}}\xspace}
\newcommand{\meanpt}       {$\langle p_{\mathrm{T}}\rangle$\xspace}
\newcommand{\ycms}         {\ensuremath{y_{\rm CMS}}\xspace}
\newcommand{\ylab}         {\ensuremath{y_{\rm lab}}\xspace}
\newcommand{\etarange}[1]  {\mbox{$\left | \eta \right |~<~#1$}}
\newcommand{\yrange}[1]    {\mbox{$\left | y \right |~<~#1$}}
\newcommand{\dndy}         {\ensuremath{\mathrm{d}N_\mathrm{ch}/\mathrm{d}y}\xspace}
\newcommand{\dndeta}       {\ensuremath{\mathrm{d}N_\mathrm{ch}/\mathrm{d}\eta}\xspace}
\newcommand{\avdndeta}     {\ensuremath{\langle\dndeta\rangle}\xspace}
\newcommand{\dNdy}         {\ensuremath{\mathrm{d}N_\mathrm{ch}/\mathrm{d}y}\xspace}
\newcommand{\Npart}        {\ensuremath{N_\mathrm{part}}\xspace}
\newcommand{\Ncoll}        {\ensuremath{N_\mathrm{coll}}\xspace}
\newcommand{\dEdx}         {\ensuremath{\textrm{d}E/\textrm{d}x}\xspace}
\newcommand{\RpPb}         {\ensuremath{R_{\rm pPb}}\xspace}

% 3) ENERGIES, UNITS
\newcommand{\nineH}           {$\sqrt{s}~=~0.9$~Te\kern-.1emV\xspace}
\newcommand{\seven}           {$\sqrt{s}~=~7$~Te\kern-.1emV\xspace}
\newcommand{\eight}            {$\sqrt{s}~=~8$~Te\kern-.1emV\xspace}
\newcommand{\thirteen}        {$\sqrt{s}~=~13$~Te\kern-.1emV\xspace}
\newcommand{\thirteensix}    {$\sqrt{s}~=~13.6$~Te\kern-.1emV\xspace}
\newcommand{\fourteen}        {$\sqrt{s}~=~14$~Te\kern-.1emV\xspace}
\newcommand{\twoH}            {$\sqrt{s}~=~0.2$~Te\kern-.1emV\xspace}
\newcommand{\twosevensix}  {$\sqrt{s}~=~2.76$~Te\kern-.1emV\xspace}
\newcommand{\five}               {$\sqrt{s}~=~5.02$~Te\kern-.1emV\xspace}
\newcommand{\fivethirtysix}   {$\sqrt{s}~=~5.36$~Te\kern-.1emV\xspace}
\newcommand{\fivefive}         {$\sqrt{s}~=~5.5$~Te\kern-.1emV\xspace}
\newcommand{\twosevensixnn}{$\sqrt{s_{\mathrm{NN}}}~=~2.76$~Te\kern-.1emV\xspace}
\newcommand{\fivenn}           {$\sqrt{s_{\mathrm{NN}}}~=~5.02$~Te\kern-.1emV\xspace}
\newcommand{\fivethirtysixnn}   {$\sqrt{s_{\mathrm{NN}}}~=~5.36$~Te\kern-.1emV\xspace}
\newcommand{\fivefivenn}         {$\sqrt{s_{\mathrm{NN}}}~=~5.5$~Te\kern-.1emV\xspace}
\newcommand{\LT}           {L{\'e}vy-Tsallis\xspace}
\newcommand{\GeVc}         {Ge\kern-.1emV/$c$\xspace}
\newcommand{\MeVc}         {Me\kern-.1emV/$c$\xspace}
\newcommand{\TeV}          {Te\kern-.1emV\xspace}
\newcommand{\GeV}          {Ge\kern-.1emV\xspace}
\newcommand{\MeV}          {Me\kern-.1emV\xspace}
\newcommand{\tev}          {Te\kern-.1emV\xspace}
\newcommand{\gev}          {\rm Ge\kern-.1emV\xspace}
\newcommand{\mev}          {\rm Me\kern-.1emV\xspace}
\newcommand{\GeVmass}      {Ge\kern-.1emV/$c^2$\xspace}
\newcommand{\MeVmass}      {Me\kern-.1emV/$c^2$\xspace}
\newcommand{\lumi}         {\ensuremath{\mathcal{L}}\xspace}

% 4) DETECTORS 
\newcommand{\ITS}          {\rm{ITS}\xspace}
\newcommand{\TOF}          {\rm{TOF}\xspace}
\newcommand{\ZDC}          {\rm{ZDC}\xspace}
\newcommand{\ZDCs}         {\rm{ZDCs}\xspace}
\newcommand{\ZNA}          {\rm{ZNA}\xspace}
\newcommand{\ZNC}          {\rm{ZNC}\xspace}
\newcommand{\SPD}          {\rm{SPD}\xspace}
\newcommand{\SDD}          {\rm{SDD}\xspace}
\newcommand{\SSD}          {\rm{SSD}\xspace}
\newcommand{\TPC}          {\rm{TPC}\xspace}
\newcommand{\TRD}          {\rm{TRD}\xspace}
\newcommand{\VZERO}        {\rm{V0}\xspace}
\newcommand{\VZEROA}       {\rm{V0A}\xspace}
\newcommand{\VZEROC}       {\rm{V0C}\xspace}
\newcommand{\Vdecay} 	   {\ensuremath{V^{0}}\xspace}

% 4) PARTICLE SPECIES 
\newcommand{\ee}           {\ensuremath{e^{+}e^{-}}} 
\newcommand{\pip}          {\ensuremath{\pi^{+}}\xspace}
\newcommand{\pim}          {\ensuremath{\pi^{-}}\xspace}
\newcommand{\kap}          {\ensuremath{\rm{K}^{+}}\xspace}
\newcommand{\kam}          {\ensuremath{\rm{K}^{-}}\xspace}
\newcommand{\pbar}         {\ensuremath{\rm\overline{p}}\xspace}
\newcommand{\kzero}        {\ensuremath{{\rm K}^{0}_{\rm{S}}}\xspace}
\newcommand{\Lmb}          {\ensuremath{\Lambda}\xspace}
\newcommand{\almb}         {\ensuremath{\overline{\Lambda}}\xspace}
\newcommand{\Om}           {\ensuremath{\Omega^-}\xspace}
\newcommand{\Mo}           {\ensuremath{\overline{\Omega}^+}\xspace}
\newcommand{\X}            {\ensuremath{\Xi^-}\xspace}
\newcommand{\Ix}           {\ensuremath{\overline{\Xi}^+}\xspace}
\newcommand{\Xis}          {\ensuremath{\Xi^{\pm}}\xspace}
\newcommand{\Oms}          {\ensuremath{\Omega^{\pm}}\xspace}
\newcommand{\degree}       {\ensuremath{^{\rm o}}\xspace}
\newcommand{\Dzero}     {\ensuremath{\rm D}^{0}\xspace}
\newcommand{\Dplus}     {\ensuremath{\rm D}^{+}\xspace}
\newcommand{\Dstar}     {\ensuremath{\rm D}^{*+}\xspace}
\newcommand{\Dstarwide} {\ensuremath{\rm D}^{*}(2010)^{+}\xspace}
\newcommand{\Lc}        {\ensuremath{\rm \Lambda}_{\rm c}^{+}\xspace}
\newcommand{\Sc}        {\ensuremath{\rm \Sigma}_{\rm c}^{0,++}\xspace}

% 5) my commands
\newcommand{\lmb}          {\ensuremath{\lambda}\xspace}
\newcommand{\R}            {\ensuremath{R}\xspace}
\newcommand{\bfivenn}{\mathbf{\sqrt{\textbf{\textit{s}}_{\textbf{NN}}} = 5.02~\textbf{TeV}}}  % bold sqrt
\newcommand{\bkzero}        {\mathbf{\ensuremath{{\textbf{K}}^{0}_{\textbf{S}}}\xspace}}

%% file: fa_2026-03-26_Opt_C.tex
% Version: 2026-03-26

The ALICE Collaboration would like to thank all its engineers and technicians for their invaluable contributions to the construction of the experiment and the CERN accelerator teams for the outstanding performance of the LHC complex.
The ALICE Collaboration gratefully acknowledges the resources and support provided by all Grid centres and the Worldwide LHC Computing Grid (WLCG) collaboration.
The ALICE Collaboration acknowledges the following funding agencies for their support in building and running the ALICE detector:
A. I. Alikhanyan National Science Laboratory (Yerevan Physics Institute) Foundation (ANSL), State Committee of Science and World Federation of Scientists (WFS), Armenia;
Austrian Academy of Sciences, Austrian Science Fund (FWF): [M 2467-N36] and Nationalstiftung f\"{u}r Forschung, Technologie und Entwicklung, Austria;
Ministry of Communications and High Technologies, National Nuclear Research Center, Azerbaijan;
Rede Nacional de Física de Altas Energias (Renafae), Financiadora de Estudos e Projetos (Finep), Funda\c{c}\~{a}o de Amparo \`{a} Pesquisa do Estado de S\~{a}o Paulo (FAPESP) and The Sao Paulo Research Foundation  (FAPESP), Brazil;
Bulgarian Ministry of Education and Science, within the National Roadmap for Research Infrastructures 2020-2027 (object CERN), Bulgaria;
Ministry of Education of China (MOEC) , Ministry of Science \& Technology of China (MSTC) and National Natural Science Foundation of China (NSFC), China;
Ministry of Science and Education and Croatian Science Foundation, Croatia;
Centro de Aplicaciones Tecnol\'{o}gicas y Desarrollo Nuclear (CEADEN), Cubaenerg\'{\i}a, Cuba;
Ministry of Education, Youth and Sports of the Czech Republic, Czech Republic;
The Danish Council for Independent Research | Natural Sciences, the VILLUM FONDEN and Danish National Research Foundation (DNRF), Denmark;
Helsinki Institute of Physics (HIP), Finland;
Commissariat \`{a} l'Energie Atomique (CEA) and Institut National de Physique Nucl\'{e}aire et de Physique des Particules (IN2P3) and Centre National de la Recherche Scientifique (CNRS), France;
Bundesministerium f\"{u}r Forschung, Technologie und Raumfahrt (BMFTR) and GSI Helmholtzzentrum f\"{u}r Schwerionenforschung GmbH, Germany;
National Research, Development and Innovation Office, Hungary;
Department of Atomic Energy Government of India (DAE), Department of Science and Technology, Government of India (DST), University Grants Commission, Government of India (UGC) and Council of Scientific and Industrial Research (CSIR), India;
National Research and Innovation Agency - BRIN, Indonesia;
Istituto Nazionale di Fisica Nucleare (INFN), Italy;
Japanese Ministry of Education, Culture, Sports, Science and Technology (MEXT) and Japan Society for the Promotion of Science (JSPS) KAKENHI, Japan;
Consejo Nacional de Ciencia (CONACYT) y Tecnolog\'{i}a, through Fondo de Cooperaci\'{o}n Internacional en Ciencia y Tecnolog\'{i}a (FONCICYT) and Direcci\'{o}n General de Asuntos del Personal Academico (DGAPA), Mexico;
Nederlandse Organisatie voor Wetenschappelijk Onderzoek (NWO), Netherlands;
The Research Council of Norway, Norway;
Pontificia Universidad Cat\'{o}lica del Per\'{u}, Peru;
Ministry of Science and Higher Education, National Science Centre and WUT ID-UB, Poland;
Korea Institute of Science and Technology Information and National Research Foundation of Korea (NRF), Republic of Korea;
Ministry of Education and Scientific Research, Institute of Atomic Physics, Ministry of Research and Innovation and Institute of Atomic Physics and Universitatea Nationala de Stiinta si Tehnologie Politehnica Bucuresti, Romania;
Ministerstvo skolstva, vyskumu, vyvoja a mladeze SR, Slovakia;
National Research Foundation of South Africa, South Africa;
Swedish Research Council (VR) and Knut \& Alice Wallenberg Foundation (KAW), Sweden;
European Organization for Nuclear Research, Switzerland;
Suranaree University of Technology (SUT), National Science and Technology Development Agency (NSTDA) and National Science, Research and Innovation Fund (NSRF via PMU-B B05F650021), Thailand;
Turkish Energy, Nuclear and Mineral Research Agency (TENMAK), Turkey;
National Academy of  Sciences of Ukraine, Ukraine;
Science and Technology Facilities Council (STFC), United Kingdom;
National Science Foundation of the United States of America (NSF) and United States Department of Energy, Office of Nuclear Physics (DOE NP), United States of America.
In addition, individual groups or members have received support from:
FORTE project, reg.\ no.\ CZ.02.01.01/00/22\_008/0004632, Czech Republic, co-funded by the European Union, Czech Republic;
European Research Council (grant no. 950692), European Union;
Deutsche Forschungs Gemeinschaft (DFG, German Research Foundation) ``Neutrinos and Dark Matter in Astro- and Particle Physics'' (grant no. SFB 1258), Germany.

%% file: CFappendix.tex
\section{Correlation function fits}
\label{app:fits}
\begin{figure}[!htbp]
    \centering
    \includegraphics[width=0.95\linewidth]{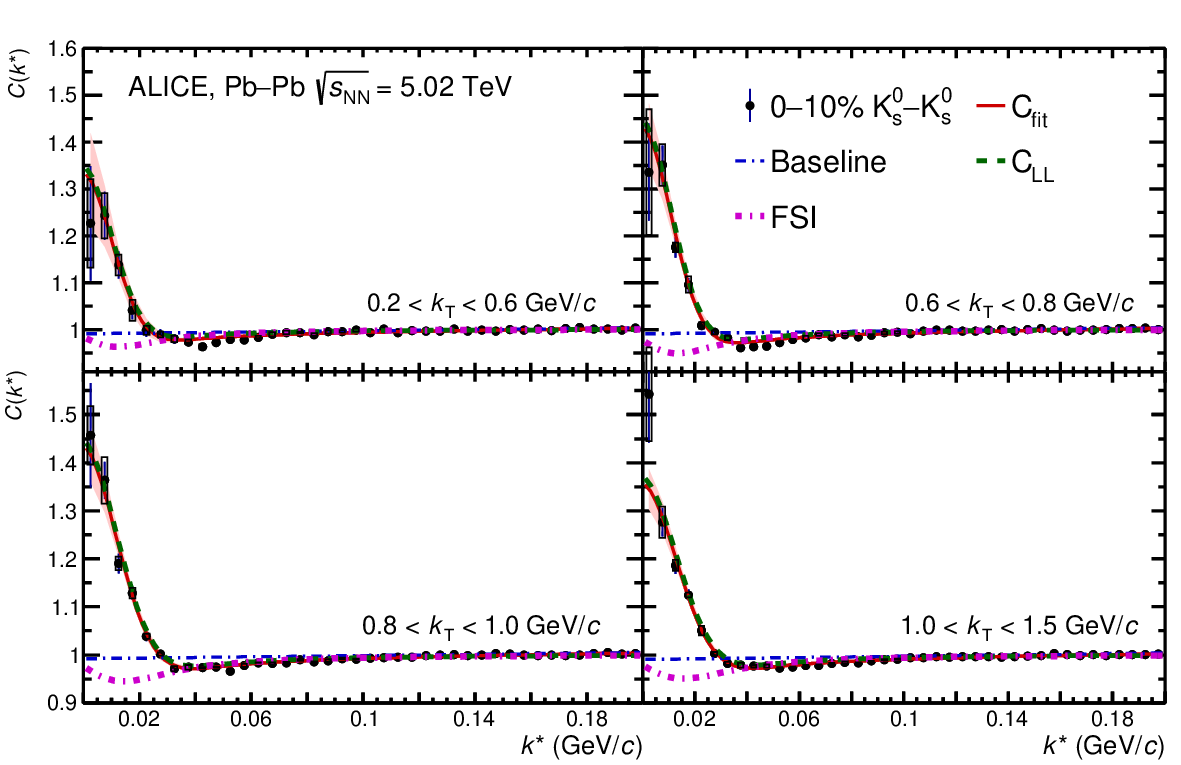}
    \caption{Correlation functions and their fits for the 0--10\% centrality range. Statistical errors are represented by bars and systematic errors by boxes. The individual contributions to the overall fit are shown as different lines with statistical uncertainty shown as the band around the fit.}
    \label{fig:CF0to10}
\end{figure}

\begin{figure}[!htbp]
    \centering
    \includegraphics[width=0.95\linewidth]{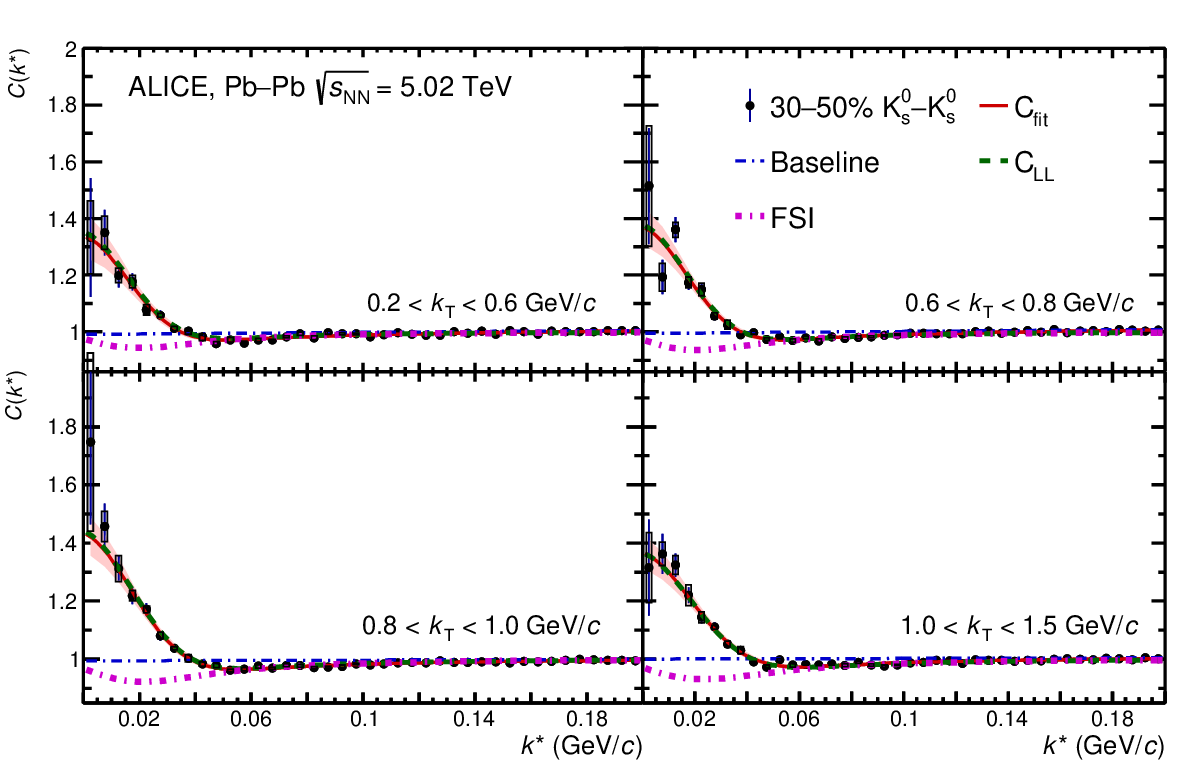}
    \caption{Correlation functions and their fits for the 30--50\% centrality range. Statistical errors are represented by bars and systematic errors by boxes. The individual contributions to the overall fit are shown as different lines with statistical uncertainty shown as the band around the fit.}
    \label{fig:CF30to50}
\end{figure}

\begin{figure}[!htbp]
    \centering
    \includegraphics[width=0.95\linewidth]{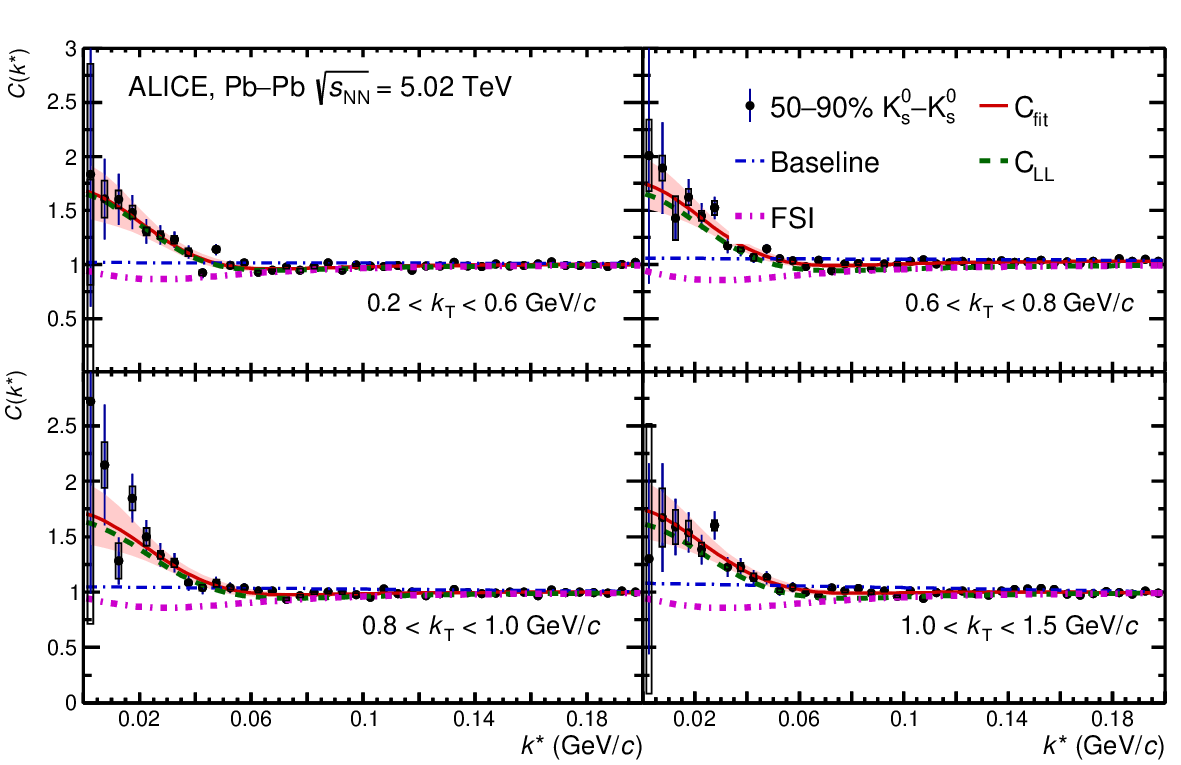}
    \caption{Correlation functions and their fits for the 50--90\% centrality range. Statistical errors are represented by bars and systematic errors by boxes. The individual contributions to the overall fit are shown as different lines with statistical uncertainty shown as the band around the fit.}
    \label{fig:CF50to90}
\end{figure}

The correlation functions and respective fits for the 0--10\%, 30--50\%, and 50--90\% centrality ranges are shown in Fig.~\ref{fig:CF0to10},~\ref{fig:CF30to50}, and~\ref{fig:CF50to90}.
A summary of all fit parameters is shown in Table~\ref{tab:LSys}.

\begin{table}[!htbp]
     \centering
     \caption{Fit parameters and fit quality for all correlation functions are reported along with statistical errors. Here, “Cent.” refers to the collision centrality, and “N” denotes the normalization parameter of the fit function. The ranges of \kt are given in \GeVc}
     \small
         \begin{tabular}{ccccccccc}
         \hline
         Cent. & \kt & $\chi ^2$/ndf & N & $R$ (fm) & $\lambda$ & $b$ & $c$ \\
         \hline
         \multirow{4}{4em}{0--10\%} & 0.2--0.6 & 0.96 & 1.08 $\pm$ 0.01 & 6.4 $\pm$ 0.5 & 0.51 $\pm$ 0.08 & 0.06 $\pm$ 0.01 & 0.653 $\pm$ 0.006  \\
            & 0.6--0.8 & 1.72 & 1.09 $\pm$ 0.01 & 5.9 $\pm$ 0.2 & 0.65 $\pm$ 0.05 & 0.058 $\pm$ 0.009 & 0.657 $\pm$ 0.004 \\
            &0.8--1.0 & 1.25 & 1.09 $\pm$ 0.01 & 5.4 $\pm$ 0.2 & 0.65 $\pm$ 0.04 & 0.058 $\pm$ 0.008 & 0.6596 $\pm$ 0.0007  \\
            & 1.0--1.5 & 1.31 & 1.09 $\pm$ 0.01 & 5.0 $\pm$ 0.1 & 0.54 $\pm$ 0.03 & 0.060 $\pm$ 0.007 & 0.6605 $\pm$ 0.0006  \\ [1em]
    
        \multirow{4}{4em}{10--30\%} & 0.2--0.6 & 1.15 & 1.01 $\pm$ 0.01 & 5.3 $\pm$ 0.2 & 0.65 $\pm$ 0.06 & 0.05 $\pm$ 0.01 & 0.605 $\pm$ 0.005 \\
            & 0.6--0.8 & 1.61 & 1.01 $\pm$ 0.01 & 4.9 $\pm$ 0.1 & 0.66 $\pm$ 0.04 & 0.062 $\pm$ 0.008 & 0.608 $\pm$ 0.004 \\
            &0.8--1.0 & 1.36 & 1.02 $\pm$ 0.01 & 4.4 $\pm$ 0.1 & 0.60 $\pm$ 0.03 & 0.047 $\pm$ 0.007 & 0.612 $\pm$ 0.003 \\
            & 1.0--1.5 & 1.41 & 1.03 $\pm$ 0.01 & 4.2 $\pm$ 0.1 & 0.54 $\pm$ 0.02 & 0.050 $\pm$ 0.006 & 0.617 $\pm$ 0.003 \\[1em]
        
        \multirow{4}{4em}{30--50\%} & 0.2--0.6 & 1.36 & 1.03 $\pm$ 0.01 & 4.0 $\pm$ 0.2 & 0.59 $\pm$ 0.06 & 0.07 $\pm$ 0.02 & 0.618 $\pm$ 0.008 \\
            & 0.6--0.8 & 1.33 & 1.04 $\pm$ 0.01 & 3.7 $\pm$ 0.1 & 0.62 $\pm$ 0.05 & 0.07 $\pm$ 0.02 & 0.624 $\pm$ 0.007\\
            &0.8--1.0 & 0.73 & 1.05 $\pm$ 0.01 & 3.6 $\pm$ 0.1 & 0.70 $\pm$ 0.05 & 0.02 $\pm$ 0.02 & 0.633 $\pm$ 0.007 \\
            & 1.0--1.5 & 1.32 & 1.07 $\pm$ 0.01 & 3.3 $\pm$ 0.1 & 0.57 $\pm$ 0.04 & 0.02 $\pm$ 0.01 & 0.642 $\pm$ 0.006 \\[1em]
        \
        \multirow{4}{4em}{50--90\%} & 0.2--0.6 & 1.19 & 2.61 $\pm$ 0.09 & 2.9 $\pm$ 0.3 & 0.60 $\pm$ 0.10 & -0.02 $\pm$ 0.03 & 0.46 $\pm$ 0.02 \\
            & 0.6--0.8 & 1.08 & 2.68 $\pm$ 0.09 & 2.6 $\pm$ 0.2 & 0.60 $\pm$ 0.09 & -0.04 $\pm$ 0.03 & 0.47 $\pm$ 0.01 \\
            &0.8--1.0 & 0.84 & 2.76 $\pm$ 0.09 & 2.6 $\pm$ 0.3 & 0.50 $\pm$ 0.10 & -0.08 $\pm$ 0.04 & 0.48 $\pm$ 0.02 \\
            & 1.0--1.5 & 1.08 & 2.88 $\pm$ 0.09 & 2.5 $\pm$ 0.2 & 0.51 $\pm$ 0.09 & -0.13 $\pm$ 0.03 & 0.50 $\pm$ 0.02 \\
        \hline
        \end{tabular}
     \label{tab:LSys}
\end{table}

%% file: Alice_Authorlist_2026-03-26_Opt_C.tex
% ALICE Collaboration author list for 2026-03-26
\begin{flushleft} 
\small

D.A.H.~Abdallah\,\orcidlink{0000-0003-4768-2718}\,$^{\rm 134}$, 
I.J.~Abualrob\,\orcidlink{0009-0005-3519-5631}\,$^{\rm 112}$, 
S.~Acharya\,\orcidlink{0000-0002-9213-5329}\,$^{\rm 49}$, 
K.~Agarwal\,\orcidlink{0000-0001-5781-3393}\,$^{\rm II,}$$^{\rm 23}$, 
G.~Aglieri Rinella\,\orcidlink{0000-0002-9611-3696}\,$^{\rm 32}$, 
L.~Aglietta\,\orcidlink{0009-0003-0763-6802}\,$^{\rm 24}$, 
N.~Agrawal\,\orcidlink{0000-0003-0348-9836}\,$^{\rm 25}$, 
Z.~Ahammed\,\orcidlink{0000-0001-5241-7412}\,$^{\rm 132}$, 
S.~Ahmad\,\orcidlink{0000-0003-0497-5705}\,$^{\rm 15}$, 
I.~Ahuja\,\orcidlink{0000-0002-4417-1392}\,$^{\rm 36}$
% ...

\section*{Affiliation Notes}

$^{\rm I}$ Deceased\\
$^{\rm II}$ Also at: INFN Trieste\\
$^{\rm III}$ Also at: Fondazione Bruno Kessler (FBK), Trento, Italy\\
$^{\rm IV}$ Also at: Czech Technical University in Prague (CZ)\\
$^{\rm V}$ Also at: Instituto de Fisica da Universidade de Sao Paulo\\
$^{\rm VI}$ Also at: Dipartimento DET del Politecnico di Torino, Turin, Italy\\
$^{\rm VII}$ Also at: Department of Applied Physics, Aligarh Muslim University, Aligarh, India\\
$^{\rm VIII}$ Also at: Institute of Theoretical Physics, University of Wroclaw, Poland\\
$^{\rm IX}$ Also at: Facultad de Ciencias, Universidad Nacional Aut\'{o}noma de M\'{e}xico, Mexico City, Mexico\\

\section*{Collaboration Institutes}

$^{1}$ A.I. Alikhanyan National Science Laboratory (Yerevan Physics Institute) Foundation, Yerevan, Armenia\\
$^{2}$ AGH University of Krakow, Cracow, Poland\\
$^{3}$ Bogolyubov Institute for Theoretical Physics, National Academy of Sciences of Ukraine, Kyiv, Ukraine\\
$^{4}$ Bose Institute, Department of Physics  and Centre for Astroparticle Physics and Space Science (CAPSS), Kolkata, India\\
$^{5}$ California Polytechnic State University, San Luis Obispo, California, United States\\
$^{6}$ Central China Normal University, Wuhan, China\\
$^{7}$ Centro de Aplicaciones Tecnol\'{o}gicas y Desarrollo Nuclear (CEADEN), Havana, Cuba\\
$^{8}$ Centro de Investigaci\'{o}n y de Estudios Avanzados (CINVESTAV), Mexico City and M\'{e}rida, Mexico\\
$^{9}$ Chicago State University, Chicago, Illinois, United States\\
$^{10}$ China Nuclear Data Center, China Institute of Atomic Energy, Beijing, China\\
$^{11}$ China University of Geosciences, Wuhan, China\\
$^{12}$ Chungbuk National University, Cheongju, Republic of Korea\\
$^{13}$ Comenius University Bratislava, Faculty of Mathematics, Physics and Informatics, Bratislava, Slovak Republic\\
$^{14}$ Creighton University, Omaha, Nebraska, United States\\
$^{15}$ Department of Physics, Aligarh Muslim University, Aligarh, India\\
$^{16}$ Department of Physics, Pusan National University, Pusan, Republic of Korea\\
$^{17}$ Department of Physics, Sejong University, Seoul, Republic of Korea\\
$^{18}$ Department of Physics, University of California, Berkeley, California, United States\\
$^{19}$ Department of Physics, University of Oslo, Oslo, Norway\\
$^{20}$ Department of Physics and Technology, University of Bergen, Bergen, Norway\\
$^{21}$ Dipartimento di Fisica, Universit\`{a} di Pavia, Pavia, Italy\\
$^{22}$ Dipartimento di Fisica dell'Universit\`{a} and Sezione INFN, Cagliari, Italy\\
$^{23}$ Dipartimento di Fisica dell'Universit\`{a} and Sezione INFN, Trieste, Italy\\
$^{24}$ Dipartimento di Fisica dell'Universit\`{a} and Sezione INFN, Turin, Italy\\
$^{25}$ Dipartimento di Fisica e Astronomia dell'Universit\`{a} and Sezione INFN, Bologna, Italy\\
$^{26}$ Dipartimento di Fisica e Astronomia dell'Universit\`{a} and Sezione INFN, Catania, Italy\\
$^{27}$ Dipartimento di Fisica e Astronomia dell'Universit\`{a} and Sezione INFN, Padova, Italy\\
$^{28}$ Dipartimento di Fisica `E.R.~Caianiello' dell'Universit\`{a} and Gruppo Collegato INFN, Salerno, Italy\\
$^{29}$ Dipartimento DISAT del Politecnico and Sezione INFN, Turin, Italy\\
$^{30}$ Dipartimento di Scienze MIFT, Universit\`{a} di Messina, Messina, Italy\\
$^{31}$ Dipartimento Interateneo di Fisica `M.~Merlin' and Sezione INFN, Bari, Italy\\
$^{32}$ European Organization for Nuclear Research (CERN), Geneva, Switzerland\\
$^{33}$ Faculty of Electrical Engineering, Mechanical Engineering and Naval Architecture, University of Split, Split, Croatia\\
$^{34}$ Faculty of Nuclear Sciences and Physical Engineering, Czech Technical University in Prague, Prague, Czech Republic\\
$^{35}$ Faculty of Physics, Sofia University, Sofia, Bulgaria\\
$^{36}$ Faculty of Science, P.J.~\v{S}af\'{a}rik University, Ko\v{s}ice, Slovak Republic\\
$^{37}$ Faculty of Technology, Environmental and Social Sciences, Bergen, Norway\\
$^{38}$ Frankfurt Institute for Advanced Studies, Johann Wolfgang Goethe-Universit\"{a}t Frankfurt, Frankfurt, Germany\\
$^{39}$ Fudan University, Shanghai, China\\
$^{40}$ Gauhati University, Department of Physics, Guwahati, India\\
$^{41}$ Helmholtz-Institut f\"{u}r Strahlen- und Kernphysik, Rheinische Friedrich-Wilhelms-Universit\"{a}t Bonn, Bonn, Germany\\
$^{42}$ Helsinki Institute of Physics (HIP), Helsinki, Finland\\
$^{43}$ High Energy Physics Group,  Universidad Aut\'{o}noma de Puebla, Puebla, Mexico\\
$^{44}$ Horia Hulubei National Institute of Physics and Nuclear Engineering, Bucharest, Romania\\
$^{45}$ HUN-REN Wigner Research Centre for Physics, Budapest, Hungary\\
$^{46}$ Indian Institute of Technology Bombay (IIT), Mumbai, India\\
$^{47}$ Indian Institute of Technology Indore, Indore, India\\
$^{48}$ INFN, Laboratori Nazionali di Frascati, Frascati, Italy\\
$^{49}$ INFN, Sezione di Bari, Bari, Italy\\
$^{50}$ INFN, Sezione di Bologna, Bologna, Italy\\
$^{51}$ INFN, Sezione di Cagliari, Cagliari, Italy\\
$^{52}$ INFN, Sezione di Catania, Catania, Italy\\
$^{53}$ INFN, Sezione di Padova, Padova, Italy\\
$^{54}$ INFN, Sezione di Pavia, Pavia, Italy\\
$^{55}$ INFN, Sezione di Torino, Turin, Italy\\
$^{56}$ INFN, Sezione di Trieste, Trieste, Italy\\
$^{57}$ Inha University, Incheon, Republic of Korea\\
$^{58}$ Institute for Gravitational and Subatomic Physics (GRASP), Utrecht University/Nikhef, Utrecht, Netherlands\\
$^{59}$ Institute of Experimental Physics, Slovak Academy of Sciences, Ko\v{s}ice, Slovak Republic\\
$^{60}$ Institute of Physics, Homi Bhabha National Institute, Bhubaneswar, India\\
$^{61}$ Institute of Physics of the Czech Academy of Sciences, Prague, Czech Republic\\
$^{62}$ Institute of Space Science (ISS), Bucharest, Romania\\
$^{63}$ Institut f\"{u}r Kernphysik, Johann Wolfgang Goethe-Universit\"{a}t Frankfurt, Frankfurt, Germany\\
$^{64}$ Instituto de Ciencias Nucleares, Universidad Nacional Aut\'{o}noma de M\'{e}xico, Mexico City, Mexico\\
$^{65}$ Instituto de F\'{i}sica, Universidade Federal do Rio Grande do Sul (UFRGS), Porto Alegre, Brazil\\
$^{66}$ Instituto de F\'{\i}sica, Universidad Nacional Aut\'{o}noma de M\'{e}xico, Mexico City, Mexico\\
$^{67}$ iThemba LABS, National Research Foundation, Somerset West, South Africa\\
$^{68}$ Jeonbuk National University, Jeonju, Republic of Korea\\
$^{69}$ Korea Institute of Science and Technology Information, Daejeon, Republic of Korea\\
$^{70}$ Laboratoire de Physique Subatomique et de Cosmologie, Universit\'{e} Grenoble-Alpes, CNRS-IN2P3, Grenoble, France\\
$^{71}$ Lawrence Berkeley National Laboratory, Berkeley, California, United States\\
$^{72}$ Lund University Department of Physics, Division of Particle Physics, Lund, Sweden\\
$^{73}$ Marietta Blau Institute, Vienna, Austria\\
$^{74}$ Nagasaki Institute of Applied Science, Nagasaki, Japan\\
$^{75}$ Nara Women{'}s University (NWU), Nara, Japan\\
$^{76}$ National Centre for Nuclear Research, Warsaw, Poland\\
$^{77}$ National Institute of Science Education and Research, Homi Bhabha National Institute, Jatni, India\\
$^{78}$ National Nuclear Research Center, Baku, Azerbaijan\\
$^{79}$ National Research and Innovation Agency - BRIN, Jakarta, Indonesia\\
$^{80}$ Niels Bohr Institute, University of Copenhagen, Copenhagen, Denmark\\
$^{81}$ Nikhef, National institute for subatomic physics, Amsterdam, Netherlands\\
$^{82}$ Nuclear Physics Group, STFC Daresbury Laboratory, Daresbury, United Kingdom\\
$^{83}$ Nuclear Physics Institute of the Czech Academy of Sciences, Husinec-\v{R}e\v{z}, Czech Republic\\
$^{84}$ Oak Ridge National Laboratory, Oak Ridge, Tennessee, United States\\
$^{85}$ Ohio State University, Columbus, Ohio, United States\\
$^{86}$ Physics department, Faculty of science, University of Zagreb, Zagreb, Croatia\\
$^{87}$ Physics Department, Panjab University, Chandigarh, India\\
$^{88}$ Physics Department, University of Jammu, Jammu, India\\
$^{89}$ Physics Program and International Institute for Sustainability with Knotted Chiral Meta Matter (WPI-SKCM$^{2}$), Hiroshima University, Hiroshima, Japan\\
$^{90}$ Physikalisches Institut, Eberhard-Karls-Universit\"{a}t T\"{u}bingen, T\"{u}bingen, Germany\\
$^{91}$ Physikalisches Institut, Ruprecht-Karls-Universit\"{a}t Heidelberg, Heidelberg, Germany\\
$^{92}$ Physik Department, Technische Universit\"{a}t M\"{u}nchen, Munich, Germany\\
$^{93}$ Politecnico di Bari and Sezione INFN, Bari, Italy\\
$^{94}$ Research Division and ExtreMe Matter Institute EMMI, GSI Helmholtzzentrum f\"ur Schwerionenforschung GmbH, Darmstadt, Germany\\
$^{95}$ Saga University, Saga, Japan\\
$^{96}$ Saha Institute of Nuclear Physics, Homi Bhabha National Institute, Kolkata, India\\
$^{97}$ School of Physics and Astronomy, University of Birmingham, Birmingham, United Kingdom\\
$^{98}$ Secci\'{o}n F\'{\i}sica, Departamento de Ciencias, Pontificia Universidad Cat\'{o}lica del Per\'{u}, Lima, Peru\\
$^{99}$ SUBATECH, IMT Atlantique, Nantes Universit\'{e}, CNRS-IN2P3, Nantes, France\\
$^{100}$ Sungkyunkwan University, Suwon City, Republic of Korea\\
$^{101}$ Suranaree University of Technology, Nakhon Ratchasima, Thailand\\
$^{102}$ Technical University of Ko\v{s}ice, Ko\v{s}ice, Slovak Republic\\
$^{103}$ The Henryk Niewodniczanski Institute of Nuclear Physics, Polish Academy of Sciences, Cracow, Poland\\
$^{104}$ The University of Texas at Austin, Austin, Texas, United States\\
$^{105}$ Universidad Aut\'{o}noma de Sinaloa, Culiac\'{a}n, Mexico\\
$^{106}$ Universidade de S\~{a}o Paulo (USP), S\~{a}o Paulo, Brazil\\
$^{107}$ Universidade Estadual de Campinas (UNICAMP), Campinas, Brazil\\
$^{108}$ Universidade Federal do ABC, Santo Andre, Brazil\\
$^{109}$ Universitatea Nationala de Stiinta si Tehnologie Politehnica Bucuresti, Bucharest, Romania\\
$^{110}$ University of Cape Town, Cape Town, South Africa\\
$^{111}$ University of Derby, Derby, United Kingdom\\
$^{112}$ University of Houston, Houston, Texas, United States\\
$^{113}$ University of Jyv\"{a}skyl\"{a}, Jyv\"{a}skyl\"{a}, Finland\\
$^{114}$ University of Kansas, Lawrence, Kansas, United States\\
$^{115}$ University of Liverpool, Liverpool, United Kingdom\\
$^{116}$ University of Science and Technology of China, Hefei, China\\
$^{117}$ University of Silesia in Katowice, Katowice, Poland\\
$^{118}$ University of South-Eastern Norway, Kongsberg, Norway\\
$^{119}$ University of Tennessee, Knoxville, Tennessee, United States\\
$^{120}$ University of the Witwatersrand, Johannesburg, South Africa\\
$^{121}$ University of Tokyo, Tokyo, Japan\\
$^{122}$ University of Tsukuba, Tsukuba, Japan\\
$^{123}$ Universit\"{a}t M\"{u}nster, Institut f\"{u}r Kernphysik, M\"{u}nster, Germany\\
$^{124}$ Universit\'{e} Clermont Auvergne, CNRS/IN2P3, LPC, Clermont-Ferrand, France\\
$^{125}$ Universit\'{e} de Lyon, CNRS/IN2P3, Institut de Physique des 2 Infinis de Lyon, Lyon, France\\
$^{126}$ Universit\'{e} de Strasbourg, CNRS, IPHC UMR 7178, F-67000 Strasbourg, France, Strasbourg, France\\
$^{127}$ Universit\'{e} Paris-Saclay, Centre d'Etudes de Saclay (CEA), IRFU, D\'{e}partment de Physique Nucl\'{e}aire (DPhN), Saclay, France\\
$^{128}$ Universit\'{e}  Paris-Saclay, CNRS/IN2P3, IJCLab, Orsay, France\\
$^{129}$ Universit\`{a} degli Studi di Foggia, Foggia, Italy\\
$^{130}$ Universit\`{a} del Piemonte Orientale, Vercelli, Italy\\
$^{131}$ Universit\`{a} di Brescia, Brescia, Italy\\
$^{132}$ Variable Energy Cyclotron Centre, Homi Bhabha National Institute, Kolkata, India\\
$^{133}$ Warsaw University of Technology, Warsaw, Poland\\
$^{134}$ Wayne State University, Detroit, Michigan, United States\\
$^{135}$ Yale University, New Haven, Connecticut, United States\\
$^{136}$ Yildiz Technical University, Istanbul, Turkey\\
$^{137}$ Yonsei University, Seoul, Republic of Korea\\
$^{138}$ Affiliated with an institute formerly covered by a cooperation agreement with CERN\\
$^{139}$ Affiliated with an international laboratory covered by a cooperation agreement with CERN.\\

\end{flushleft} 